\documentclass[oneside,reqno,english]{amsart}
\usepackage[latin9]{inputenc}
\usepackage{array}
\usepackage{multirow}
\usepackage{amsbsy}
\usepackage{amstext}
\usepackage{amsthm}
\usepackage{amssymb}
\usepackage{stmaryrd}
\usepackage{graphicx}
\usepackage{geometry}
\usepackage{setspace}

\makeatletter

\providecommand{\tabularnewline}{\\}

\theoremstyle{plain}
\newtheorem{prop}{\protect\propositionname}

\@ifundefined{date}{}{\date{}}
\PassOptionsToPackage{reqno}{amsmath}

\usepackage{mathtools}

\makeatletter
\AtBeginDocument{\tagsleft@false} 
\makeatother

\usepackage{chngcntr}
\counterwithout{equation}{section}

\usepackage{microtype}   
\usepackage{setspace}    
\usepackage[table]{xcolor}

\makeatother

\usepackage{babel}
\providecommand{\propositionname}{Proposition}

\begin{document}
\title{The Zeta Tail Distribution:\\
A Novel Event-Count Model}
\author{Michael R. Powers}
\address{Department of Finance, School of Economics and Management, Tsinghua
University, Beijing, China 100084}
\email{powers@sem.tsinghua.edu.cn}
\date{June 23, 2026}
\begin{abstract}
We introduce the $\textrm{Zeta Tail}\left(a\right)$ probability distribution
as a new model for random damage-event counts in risk analysis. Although
a natural analogue of the $\textrm{Geometric}\left(p\right)$ distribution,
$\textrm{Zeta Tail}\left(a\right)$ has received little attention
in the scholarly literature. In the present work, we show this distribution
to be reasonably tractable by deriving various fundamental properties,
including moments, generating functions, and reliability functions.
We then assess its usefulness as an alternative to $\textrm{Geometric}\left(p\right)$,
both theoretically and through application to a set of meteorological
data. Finally, we discuss conceptual differences between employing
the $\textrm{Zeta Tail}\left(a\right)$ model conditionally (i.e.,
given observed data with certain known characteristics) and unconditionally
(i.e., for arbitrary, as yet unobserved data).
\end{abstract}

\maketitle

\section{Introduction}

\noindent In the present study, we introduce the $\textrm{Zeta Tail}\left(a\right)$
probability distribution as a new model for random damage-event counts
in risk analysis. The distribution is characterized by its probability
mass function (PMF),
\begin{equation}
f_{X\mid a}^{\left(\textrm{ZT}\right)}\left(x\right)=a\zeta\left(x+1,a+1\right),\quad x\in\mathbb{Z}_{\geq1},a\in\mathbb{R}_{>0},
\end{equation}
\begin{equation}
=a\left[\zeta\left(x+1,a\right)-a^{-\left(x+1\right)}\right],
\end{equation}
where
\[
\zeta\left(s,t\right)={\displaystyle \sum_{i=0}^{\infty}}\dfrac{1}{\left(i+t\right)^{s}},\quad s\in\mathbb{R}_{>1},t\in\mathbb{R}_{>0}
\]
denotes the Hurwitz zeta function. Summing the PMF in (1) yields the
cumulative distribution function (CDF)
\[
F_{X\mid a}^{\left(\textrm{ZT}\right)}\left(x\right)=a\sum_{k=1}^{x}\zeta\left(k+1,a+1\right),\quad x\in\mathbb{Z}_{\geq1},a\in\mathbb{R}_{>0},
\]
for which it is straightforward to confirm that
\[
\underset{x\rightarrow\infty}{\lim}F_{X\mid a}^{\left(\textrm{ZT}\right)}\left(x\right)=a\sum_{k=1}^{\infty}\zeta\left(k+1,a+1\right)
\]
\[
=a\sum_{k=1}^{\infty}\left[{\displaystyle \sum_{i=0}^{\infty}}\dfrac{1}{\left(i+a+1\right)^{k+1}}\right]
\]
\[
=a\sum_{i=0}^{\infty}\left[{\displaystyle \sum_{k=1}^{\infty}}\dfrac{1}{\left(i+a+1\right)^{k+1}}\right]
\]
\[
=a\sum_{i=0}^{\infty}\dfrac{\left(i+a+1\right)^{-2}}{1-\left(i+a+1\right)^{-1}}
\]
\[
=a\sum_{i=0}^{\infty}\dfrac{1}{\left(i+a\right)\left(i+a+1\right)}
\]
\[
=a\sum_{i=0}^{\infty}\left(\dfrac{1}{i+a}-\dfrac{1}{i+a+1}\right)=1.
\]

\begin{singlespace}
From (2), one can see that the $\textrm{Zeta Tail}\left(a\right)$
PMF is directly proportional to the (positive) difference between
the Hurwitz zeta function, $\zeta\left(x+1,a\right)$, and its asymptotic
order of magnitude in the tail, $a^{-\left(x+1\right)}$, thereby
motivating our choice of ``Zeta Tail'' for the distribution's name.
Although a natural analogue of the $\textrm{Geometric}\left(p\right)$
distribution, $\textrm{Zeta Tail}\left(a\right)$ has received little
attention in the scholarly literature, presumably (at least in part)
because its PMF is expressed in terms of a special mathematical function
applied to the index $x$. Whereas special functions commonly appear
among the normalizing constants of well-known probability distributions
(e.g., $\textrm{Negative Binomial}\left(r,p\right)$, $\textrm{Gamma}\left(r,\beta\right)$,
$\textrm{Beta}\left(a,b\right)$, etc.), they rarely are encountered
as functions of $x$ itself. One example of the Hurwitz zeta function
as a normalizing constant appears in the discrete $\textrm{Log Hurwitz Zeta}\left(b,d\right)$
distribution, a generalization of the $\textrm{Zeta}\left(b\right)$
distribution identified in a remark by Hu et al. (2006) and characterized
by the PMF
\[
f_{X\mid b,d}^{\left(\textrm{HZ}\right)}\left(x\right)=\dfrac{1}{\zeta\left(b+1,d\right)\left(x+d\right)^{b+1}},\quad x\in\mathbb{Z}_{\geq0},b\in\mathbb{R}_{>0},d\in\mathbb{R}_{\geq1}.
\]

Despite the central role of $\zeta\left(\cdot,\cdot\right)$ in (1)
and (2), we have found the novel $\textrm{Zeta Tail}\left(a\right)$
distribution to be reasonably tractable, and now investigate its fundamental
properties and potential applications. In doing so, it is useful to
distinguish between two forms of the distribution: the ordinary $\textrm{Zeta Tail}\left(a\right)$
model of (1); and a shifted version, $\textrm{Zeta Tail 0}\left(a\right)$,
defined on the sample space $\mathbb{Z}_{\geq0}$. We begin with the
ordinary version because of its ease of comparison with other discrete
probability distributions, such as $\textrm{Zeta}\left(b\right)$
and $\textrm{Geometric}\left(p\right)$, that often are defined on
$\mathbb{Z}_{\geq1}$. Moreover, we sometimes use the ordinary name
generically to cover both cases, as in the article's title and abstract.
However, when considering specific event-count applications likely
to involve the possibility of $0$ counts (as in Sections 4 and 5),
we will switch to working with the shifted version and make comparisons
to other distributions defined on $\mathbb{Z}_{\geq0}$ (whose names
indicate the shifted sample space by including a ``$0$''; e.g.,
$\textrm{Geometric 0}\left(p\right)$).

One important aspect of the $\textrm{Zeta Tail}\left(a\right)$ distribution
is that it can be formed as a countable mixture of $\textrm{Geometric}\left(p\right)$
models. In general, mixture distributions provide a versatile and
widely used framework for modeling random phenomena, characterized
by both: (i) a theoretical structure capable of incorporating an array
of coexisting processes differentiated by random perturbations; and
(ii) the practical flexibility to replicate empirically observed patterns
through a set of adjustable components. As noted by Powers and Xu
(2025b), these characteristics make such models especially appropriate
for the highly complex and chaotic physical processes of the geosciences
(e.g., seismology, hydrology, and meteorology), as well as the risks
to society posed by those processes (e.g., earthquakes, floods, and
cyclones).
\end{singlespace}

$\textrm{Geometric}\left(p\right)$ mixtures are particularly well
suited for modeling waiting times in the geosciences, as illustrated
by their frequent use in the study of successive wet and dry periods
in meteorology (see, e.g., Foufoula-Georgiou and Lettenmaier, 1987;
Dobi-Wantuch, Mika, and Szeidl, 2000; and Deni and Jemain, 2009).
In risk analysis, they commonly are used to model the discrete counts
of damage-causing events, such as loss frequencies in actuarial finance
(see, e.g., Hesselager, 1994; Willmot and Woo, 2013; and Chiu and
Yin, 2014).

\begin{singlespace}
After motivating the form of the $\textrm{Zeta Tail}\left(a\right)$
distribution in Section 2, we derive analytic expressions for its
moments, generating functions, reliability functions, and formulation
as a mixture distribution in Section 3. Then, the usefulness of $\textrm{Zeta Tail}\left(a\right)$
as an alternative to $\textrm{Geometric}\left(p\right)$ -- considered
in their shifted forms, $\textrm{Zeta Tail 0}\left(a\right)$ and
$\textrm{Geometric 0}\left(p\right)$, respectively -- is assessed
in Section 4, both theoretically and through application to a set
of South Korean rainfall data. Finally, Section 5 addresses conceptual
differences between employing the $\textrm{Zeta Tail 0}\left(a\right)$
model \emph{conditionally} (that is, given a set of observed data
with known characteristics) and \emph{unconditionally} (that is, for
arbitrary sets of unobserved data).
\end{singlespace}

\section{Motivation}

\begin{onehalfspace}
\noindent For the special case of $a=1$, the $\textrm{Zeta Tail}\left(a\right)$
PMF simplifies to
\begin{equation}
f_{X\mid a=1}^{\left(\textrm{ZT}\right)}\left(x\right)=\zeta\left(x+1\right)-1,\quad x\in\mathbb{Z}_{\geq1},
\end{equation}
where\newpage
\[
\zeta\left(s\right)={\displaystyle \sum_{i=0}^{\infty}}\dfrac{1}{\left(i+1\right)^{s}},\quad s\in\mathbb{R}_{>1}
\]
denotes the ordinary Riemann zeta function. This special case was
considered by Keith (2010) in the only published reference we have
found involving the $\textrm{Zeta Tail}\left(a\right)$ distribution.
Keith noted that his work was inspired by the challenge of a colleague
to provide a probabilistic interpretation of the PMF implied by the
right-hand side of (3).
\end{onehalfspace}

Interestingly, (3) emerges simply and elegantly from the remarkable
infinite-series identity
\begin{equation}
{\displaystyle \sum_{m=2}^{\infty}}\:{\displaystyle \sum_{n=2}^{\infty}}\dfrac{1}{m^{n}}=1,
\end{equation}

\noindent which reveals that all possible powers of the form $m^{n}$
(such that $m,n\in\left\{ 2,3,\ldots\right\} $) are bound together
by the sum of their inverses. As shown in Table 1, the two summations
in (4) can be performed in either order.\medskip{}

\begin{center}
Table 1. Partitioning $1$ into Inverses of $m^{n}$
\par\end{center}

\begin{center}
\begin{tabular}{c|c|c|c|c|c|c|}
\multicolumn{1}{c}{} & \multicolumn{1}{c}{} & \multicolumn{4}{c}{{\small\textbf{$\boldsymbol{n=}$}}} & \multicolumn{1}{c}{}\tabularnewline
\cline{3-7}
\multicolumn{1}{c}{} &  & {\small\textbf{$\boldsymbol{2}$}} & {\small\textbf{$\boldsymbol{3}$}} & {\small\textbf{$\boldsymbol{4}$}} & {\small\textbf{$\boldsymbol{\ldots}$}} & {\small\textbf{$\boldsymbol{\underset{}{{\displaystyle \sum_{n=2}^{\infty}}\dfrac{1}{m^{n}}}}$}}\tabularnewline
\cline{2-7}
\multirow{4}{*}{{\small\textbf{$\boldsymbol{m=}$}}} & {\small\textbf{$\boldsymbol{2}$}} & {\small\textbf{$\dfrac{1}{2^{2}}$}} & {\small\textbf{$\dfrac{1}{2^{3}}$}} & {\small\textbf{$\dfrac{1}{2^{4}}$}} & {\small\textbf{$\ldots$}} & {\small\textbf{$\underset{}{1-\dfrac{1}{2}}$}}\tabularnewline
\cline{2-7}
 & {\small\textbf{$\boldsymbol{3}$}} & {\small\textbf{$\dfrac{1}{3^{2}}$}} & {\small\textbf{$\dfrac{1}{3^{3}}$}} & {\small\textbf{$\dfrac{1}{3^{4}}$}} & {\small\textbf{$\ldots$}} & {\small\textbf{$\underset{}{\dfrac{1}{2}-\dfrac{1}{3}}$}}\tabularnewline
\cline{2-7}
 & {\small\textbf{$\boldsymbol{4}$}} & {\small\textbf{$\dfrac{1}{4^{2}}$}} & {\small\textbf{$\dfrac{1}{4^{3}}$}} & {\small\textbf{$\dfrac{1}{4^{4}}$}} & {\small\textbf{$\ldots$}} & {\small\textbf{$\underset{}{\dfrac{1}{3}-\dfrac{1}{4}}$}}\tabularnewline
\cline{2-7}
 & {\small\textbf{$\boldsymbol{\vdots}$}} & {\small\textbf{$\vdots$}} & {\small\textbf{$\vdots$}} & {\small\textbf{$\vdots$}} & {\small\textbf{$\ddots$}} & {\small\textbf{$\vdots$}}\tabularnewline
\cline{2-7}
 & {\small\textbf{$\boldsymbol{\underset{}{{\displaystyle \sum_{m=2}^{\infty}}\dfrac{1}{m^{n}}}}$}} & {\small\textbf{$\zeta\left(2\right)-1$}} & {\small\textbf{$\zeta\left(3\right)-1$}} & {\small\textbf{$\zeta\left(4\right)-1$}} & {\small\textbf{$\ldots$}} & {\small\textbf{$\underset{}{{\displaystyle \sum_{m=2}^{\infty}}\:{\displaystyle \sum_{n=2}^{\infty}}\dfrac{1}{m^{n}}=1}$}}\tabularnewline
\cline{2-7}
\end{tabular}
\par\end{center}

\begin{center}
\medskip{}
\par\end{center}

In each column $n\in\left\{ 2,3,\ldots\right\} $, the $m^{\textrm{th}}$
entry, $1/m^{n}$, is proportional to the value of the PMF of the
$\textrm{Zeta}\left(b=n-1\right)$ distribution (defined on the sample
space $\mathbb{Z}_{\geq1}$) at $x=m\in\left\{ 2,3,\ldots\right\} $;
that is,
\[
f_{X\mid b=n-1}^{\left(\textrm{Z}\right)}\left(m\right)=\left.\dfrac{1}{\zeta\left(b+1\right)x^{b+1}}\right|_{b=n-1,x=m},\quad m\in\mathbb{Z}_{\geq2}
\]
\[
=\dfrac{1}{\zeta\left(n\right)}\cdot\dfrac{1}{m^{n}}.
\]
Moreover, adding these values together across columns gives the exact
value of the $\textrm{Quadratic}\left(c=1\right)$ distribution\footnote{The name and parameterization of this distribution are given by Powers
and Xu (2025c), where the distribution is defined on the sample space
$\mathbb{Z}_{\geq0}$ (and therefore equivalent to $\textrm{Quadratic 0}\left(c=1\right)$
in our present terminology).} at $x=m-1$:
\[
f_{X\mid c=1}^{\left(\textrm{Q}\right)}\left(m-1\right)=\left.\dfrac{c}{\left(x-1+c\right)\left(x+c\right)}\right|_{c=1,x=m-1},\quad m\in\mathbb{Z}_{\geq2}
\]
\[
=\dfrac{1}{\left(m-1\right)m}.
\]
Similarly, in each row $m\in\left\{ 2,3,\ldots\right\} $, the $n^{\textrm{th}}$
entry, $1/m^{n}$, is proportional to the value of the PMF of the
$\textrm{Geometric}\left(p=\tfrac{m-1}{m}\right)$ distribution at
$x=n\in\left\{ 2,3,\ldots\right\} $; that is,
\[
f_{X\mid p=\tfrac{m-1}{m}}^{\left(\textrm{G}\right)}\left(n\right)=\left.p\left(1-p\right)^{x-1}\right|_{p=\tfrac{m-1}{m},x=n},\quad n\in\mathbb{Z}_{\geq2}
\]
\[
=\left(m-1\right)\cdot\dfrac{1}{m^{n}}.
\]
Then, when we add these values together across rows, we obtain the
exact value of the $\textrm{Zeta Tail}\left(a=1\right)$ distribution
at $x=n-1$:
\[
f_{X\mid a=1}^{\left(\textrm{ZT}\right)}\left(n-1\right)=\left.a\zeta\left(x+1,a+1\right)\right|_{a=1,x=n-1},\quad n\in\mathbb{Z}_{\geq2}
\]
\[
=\zeta\left(n\right)-1.
\]

This breakdown of the various components in Table 1 shows that the
relationship between the\linebreak{}
$\textrm{Zeta Tail}\left(a=1\right)$ and $\textrm{Geometric}\left(p=\left(m-1\right)/m\right)$
distributions is similar to that between the $\textrm{Quadratic}\left(c=1\right)$
and $\textrm{Zeta}\left(b=n-1\right)$ distributions. In fact, the
former relationship is even stronger in the sense that the $\textrm{Zeta Tail}\left(a=1\right)$
PMF can be expressed as a countable mixture of $\textrm{Geometric}\left(p=m/\left(m+1\right)\right)$
PMFs,
\begin{equation}
f_{X\mid a=1}^{\left(\textrm{ZT}\right)}\left(x\right)=\sum_{m=1}^{\infty}f_{X\mid p=\tfrac{m}{m+1}}^{\left(\textrm{G}\right)}\left(x\right)f_{X\mid c=1}^{\left(\textrm{Q}\right)}\left(m\right),
\end{equation}
whereas the $\textrm{Quadratic}\left(c=1\right)$ PMF cannot be written
as a mixture of $\textrm{Zeta}\left(b\right)$ PMFs. (The impossibility
of constructing $f_{X\mid c=1}^{\left(\textrm{Q}\right)}\left(x\right)$
as either a discrete or continuous mixture of $f_{X\mid b}^{\left(\textrm{Z}\right)}\left(x\right)$
is addressed by Proposition A.1 of the Appendix.) As will be shown
in Proposition 4 of the following section, the mixture model of (5)
can be generalized to the entire family of $\textrm{Zeta Tail}\left(a\right)$
distributions.

\section{Basic Distribution Properties}

\noindent Given that $\zeta\left(x+1,a+1\right)\sim\left(a+1\right)^{-\left(x+1\right)}$
as $x\rightarrow\infty$, the tail of the $\textrm{Zeta Tail}\left(a\right)$
PMF satisfies $f_{X\mid a}^{\left(\textrm{ZT}\right)}\left(x\right)\asymp\left(a+1\right)^{-x}$,
and thus is comparable to the $\textrm{Geometric}\left(p\right)$
tail, $f_{X\mid p}^{\left(\textrm{G}\right)}\left(x\right)\asymp\left(1-p\right)^{x}$,
for $p=a/\left(a+1\right)$. Furthermore, taking first differences
of the $\textrm{Zeta Tail}\left(a\right)$ PMF with respect to $x$
yields
\[
f_{X\mid a}^{\left(\textrm{ZT}\right)}\left(x+1\right)-f_{X\mid a}^{\left(\textrm{ZT}\right)}\left(x\right)=a\left(\zeta\left(x+2,a+1\right)-\zeta\left(x+1,a+1\right)\right)
\]
\[
=a\left[\sum_{k=0}^{\infty}\dfrac{1}{\left(k+a+1\right)^{x+2}}-\sum_{k=0}^{\infty}\dfrac{1}{\left(k+a+1\right)^{x+1}}\right]
\]
\[
=-a\sum_{k=0}^{\infty}\dfrac{k+a}{\left(k+a+1\right)^{x+2}}<0,
\]
implying that the PMF is strictly decreasing over its entire sample
space (with mode at $x=1$) -- another similarity with the $\textrm{Geometric}\left(p\right)$
PMF. 

The four propositions below summarize the new distribution's basic
properties, providing analytic expressions for its moments, generating
functions, reliability functions, and formulation as a mixture distribution.
These results make use of several special functions, including the
Hurwitz zeta function (defined in the Introduction), Stirling numbers
of the second kind,
\[
S\left(\kappa,\nu\right)=\dfrac{1}{\nu!}{\displaystyle \sum_{i=0}^{\nu}}\left(-1\right)^{i}\dbinom{\nu}{i}\left(\nu-i\right)^{\kappa},\quad\kappa,\nu\in\mathbb{Z}_{\geq0},
\]
and the digamma function,
\[
\psi\left(z\right)=\dfrac{\Gamma^{\prime}\left(z\right)}{\Gamma\left(z\right)}=-\gamma+{\displaystyle \sum_{i=0}^{\infty}\left(\dfrac{1}{i+1}-\dfrac{1}{i+z}\right)},\quad z\in\mathbb{C}\setminus\mathbb{Z}_{\leq0},
\]
where $\gamma$ denotes the Euler-Mascheroni constant. However, since
each of these functions is well known, cognitively accessible, and
easily handled by computer software and artificial-intelligence systems,
we believe $\textrm{Zeta Tail}\left(a\right)$ is sufficiently ``tractable''
for most practical applications. Along these lines, it is worth noting
that certain commonly used distributions are characterized by these
special functions; for example, the $\kappa^{\textrm{th}}$ raw moment
of $\textrm{Negative Binomial 0}\left(r,p\right)$ (with sample space
$\mathbb{Z}_{\geq0}$) is
\begin{equation}
\textrm{E}_{X\mid r,p}^{\left(\textrm{NB0}\right)}\left[X^{\kappa}\right]={\displaystyle \sum_{\nu=1}^{\kappa}S\left(\kappa,\nu\right)\dfrac{\Gamma\left(r+\nu\right)}{\Gamma\left(r\right)}\left(\dfrac{1-p}{p}\right)^{\nu}},\quad\kappa\in\mathbb{Z}_{\geq1}
\end{equation}
(see Powers and Xu, 2025c).\medskip{}

\begin{prop}
For $\kappa\in\mathbb{Z}_{>0}$, the $\kappa^{\textrm{th}}$ raw moment
of $X\mid a\sim\textrm{Zeta Tail}\left(a\right)$ is given by
\begin{equation}
\textrm{E}_{X\mid a}^{\left(\textrm{ZT}\right)}\left[X^{\kappa}\right]=a{\displaystyle \sum_{\nu=1}^{\kappa}}S\left(\kappa,\nu\right)\nu!\zeta\left(\nu+1,a\right).
\end{equation}
\end{prop}
\begin{proof}
\begin{singlespace}
\phantom{}
\end{singlespace}

\begin{singlespace}
\medskip{}
\end{singlespace}

\begin{singlespace}
\noindent See the Appendix.
\end{singlespace}
\end{proof}
Note that the $\textrm{Zeta Tail}\left(a\right)$ raw moment in (7)
is quite similar to the corresponding expression for the $\textrm{Geometric 0}\left(p\right)$
distribution (with sample space $\mathbb{Z}_{\geq0}$),
\[
\textrm{E}_{X\mid p}^{\left(\textrm{G0}\right)}\left[X^{\kappa}\right]={\displaystyle \sum_{\nu=1}^{\kappa}}S\left(\kappa,\nu\right)\nu!\left(\dfrac{1-p}{p}\right)^{\nu},
\]
which is derived by setting $r=1$ in the $\textrm{Negative Binomial 0}\left(r,p\right)$
raw moment of (6). Inserting successive values of $\kappa=1,2,\ldots$
into (7) yields
\begin{equation}
\begin{array}{c}
\textrm{E}_{X\mid a}^{\left(\textrm{ZT}\right)}\left[X\right]=a\zeta\left(2,a\right),\\
\textrm{E}_{X\mid a}^{\left(\textrm{ZT}\right)}\left[X^{2}\right]=a\left[\zeta\left(2,a\right)+2\zeta\left(3,a\right)\right],\\
\textrm{E}_{X\mid a}^{\left(\textrm{ZT}\right)}\left[X^{3}\right]=a\left[\zeta\left(2,a\right)+6\zeta\left(3,a\right)+6\zeta\left(4,a\right)\right],\\
\vdots
\end{array}
\end{equation}

\noindent revealing that the raw moments remain tractable, but increasingly
complex, as $\kappa$ increases. Naturally, one can use the $1^{\textrm{st}}$
through $\kappa^{\textrm{th}}$ raw moments to construct the $\kappa^{\textrm{th}}$
central moment, $\textrm{E}_{X\mid a}^{\left(\textrm{ZT}\right)}\left[\left(X-\textrm{E}_{X\mid a}^{\left(\textrm{ZT}\right)}\left[X\right]\right)^{\kappa}\right]$,
for all $\kappa$. However, as illustrated by the $2^{\textrm{nd}}$
central moment,
\[
\textrm{E}_{X\mid a}^{\left(\textrm{ZT}\right)}\left[\left(X-\textrm{E}_{X\mid a}^{\left(\textrm{ZT}\right)}\left[X\right]\right)^{2}\right]=\textrm{Var}_{X\mid a}^{\left(\textrm{ZT}\right)}\left[X\right]
\]
\begin{equation}
=a\left[\zeta\left(2,a\right)+2\zeta\left(3,a\right)-a\left(\zeta\left(2,a\right)\right)^{2}\right],
\end{equation}
these quantities do not possess a particularly elegant form.\medskip{}

\begin{prop}
The probability generating function, moment-generating function, Laplace
transform, and characteristic function of $X\mid a\sim\textrm{Zeta Tail}\left(a\right)$
are given respectively by:

(i) $\textrm{G}_{X\mid a}^{\left(\textrm{ZT}\right)}\left(z\right)=\textrm{E}_{X\mid a}^{\left(\textrm{ZT}\right)}\left[z^{X}\right]=a\left(\psi\left(a+1\right)-\psi\left(a+1-z\right)\right)$,
for $\left|z\right|<a+1$;

(ii) $\textrm{M}_{X\mid a}^{\left(\textrm{ZT}\right)}\left(t\right)=\textrm{E}_{X\mid a}^{\left(\textrm{ZT}\right)}\left[e^{tX}\right]=a\left(\psi\left(a+1\right)-\psi\left(a+1-e^{t}\right)\right)$,
for $t<\ln\left(a+1\right)$;

(iii) $\mathcal{L}_{X\mid a}^{\left(\textrm{ZT}\right)}\left(s\right)=\textrm{E}_{X\mid a}^{\left(\textrm{ZT}\right)}\left[e^{-sX}\right]=a\left(\psi\left(a+1\right)-\psi\left(a+1-e^{-s}\right)\right)$,
for $s>-\ln\left(a+1\right)$; and

(iv) $\varphi_{X\mid a}^{\left(\textrm{ZT}\right)}\left(\omega\right)=\textrm{E}_{X\mid a}^{\left(\textrm{ZT}\right)}\left[e^{i\omega X}\right]=a\left(\psi\left(a+1\right)-\psi\left(a+1-e^{i\omega}\right)\right)$,
for $\omega\in\mathbb{R}$.
\end{prop}
\begin{proof}
\begin{singlespace}
\phantom{}
\end{singlespace}

\begin{singlespace}
\medskip{}
\end{singlespace}

\begin{singlespace}
\noindent See the Appendix.
\end{singlespace}
\end{proof}
The expressions in (i)-(iv) can be used to facilitate construction
of a new, $2$-parameter distribution simply by adding independent
and identically distributed (IID) $\textrm{Zeta Tail}\left(a\right)$
random variables, in the same way that the $\textrm{Negative Binomial}\left(r,p\right)$
distribution (with sample space $\mathbb{Z}_{\geq r}$ for $r\in\mathbb{Z}_{\geq1}$)
is formed from $\textrm{Geometric}\left(p\right)$ random variables
(with sample space $\mathbb{Z}_{\geq1}$). Specifically, if we set
$Y={\displaystyle {\textstyle \sum_{i=1}^{r}}X_{i}},$ where the $X_{i}\mid a\overset{\textrm{IID}}{\sim}\textrm{Zeta Tail}\left(a\right)$,
then
\[
Y\mid r,a\sim\textrm{Generalized Zeta Tail}\left(r,a\right),\quad y\in\mathbb{Z}_{\geq r},r\in\mathbb{Z}_{\geq1},a\in\mathbb{R}_{>0}
\]
with generating functions derived by raising each of the quantities
in (i)-(iv) to the $r^{\textrm{th}}$ power. (For example, $\textrm{M}_{Y\mid r,a}^{\left(\textrm{GZT}\right)}\left(t\right)=\textrm{E}_{Y\mid r,a}^{\left(\textrm{GZT}\right)}\left[e^{tY}\right]=\left[a\left(\psi\left(a+1\right)-\psi\left(a+1-e^{t}\right)\right)\right]^{r}$.)\medskip{}

\begin{prop}
The hazard function and mean residual-life function of $X\mid a\sim\textrm{Zeta Tail}\left(a\right)$
are given respectively by:

(i) $h_{X\mid a}^{\left(\textrm{ZT}\right)}\left(x\right)=\dfrac{f_{X\mid a}^{\left(\textrm{ZT}\right)}\left(x\right)}{1-F_{X\mid a}^{\left(\textrm{ZT}\right)}\left(x\right)}=\dfrac{\zeta\left(x+1,a+1\right)}{{\displaystyle {\textstyle \sum}_{k=x+1}^{\infty}}\zeta\left(k+1,a+1\right)}$,
which is strictly decreasing in $x\in\mathbb{Z}_{\geq1}$\footnote{For a random variable $X\sim F_{X}\left(x\right),\:x\in\mathbb{Z}_{\geq1}$,
we define the discrete hazard function as $h_{X}\left(x\right)=f_{X}\left(x\right)/\left[1-F_{X}\left(x\right)\right]$,
the convention in actuarial finance. In some fields, it is more common
to work with $\overline{h}_{X}\left(x\right)=f_{X}\left(x\right)/\left[1-F_{X}\left(x-1\right)\right]$
(where $F_{X}\left(0\right)=0$), a quantity less than $1$ for all
$x$.} with $\underset{x\rightarrow\infty}{\lim}h_{X\mid a}^{\left(\textrm{ZT}\right)}\left(x\right)=a$;
and

(ii) $e_{X\mid a}^{\left(\textrm{ZT}\right)}\left(x\right)=\textrm{E}_{X\mid a}^{\left(\textrm{ZT}\right)}\left[\left.X-x\right|X>x\right]=\dfrac{\zeta\left(2,a\right)-x/a+{\displaystyle {\textstyle \sum}_{k=1}^{x-1}\left(x-k\right)}\zeta\left(k+1,a+1\right)}{{\displaystyle {\textstyle \sum}_{k=x+1}^{\infty}}\zeta\left(k+1,a+1\right)}$,
which is strictly increasing in $x\in\mathbb{Z}_{\geq1}$ with $\underset{x\rightarrow\infty}{\lim}e_{X\mid a}^{\left(\textrm{ZT}\right)}\left(x\right)=\dfrac{a+1}{a}$.
\end{prop}
\begin{proof}
\begin{singlespace}
\phantom{}
\end{singlespace}

\begin{singlespace}
\medskip{}
\end{singlespace}

\begin{singlespace}
\noindent See the Appendix.
\end{singlespace}
\end{proof}
\begin{prop}
The $\textrm{Zeta Tail}\left(a\right)$ PMF can be written as a countable
mixture of $\textrm{Geometric}\left(p\right)$ PMFs as follows:
\begin{equation}
f_{X\mid a}^{\left(\textrm{ZT}\right)}\left(x\right)=\sum_{m=1}^{\infty}f_{X\mid p=\tfrac{m+a-1}{m+a}}^{\left(\textrm{G}\right)}\left(x\right)f_{X\mid c=a}^{\left(\textrm{Q}\right)}\left(m\right).
\end{equation}
\end{prop}
\begin{proof}
\begin{singlespace}
\phantom{}
\end{singlespace}

\begin{singlespace}
\medskip{}
\end{singlespace}

\begin{singlespace}
\noindent See the Appendix.
\end{singlespace}
\end{proof}
The mixture formulation in (10) admits an intuitive extension of the
``number of trials needed for the first success in a Bernoulli sequence''
interpretation of the $\textrm{Geometric}\left(p\right)$ random variable.
That is, before generating the sequence of $X_{i}\mid p\overset{\textrm{IID}}{\sim}\textrm{Bernoulli}\left(p\right)$
underlying $\textrm{Geometric}\left(p\right)$, one simply chooses
the ``probability of success'' parameter ($p$) from a distinct
(and independent) random process such that 
\[
\Pr\left\{ p=\dfrac{m+a-1}{m+a}\right\} =\dfrac{a}{\left(m+a-1\right)\left(m+a\right)}
\]
for fixed $a\in\mathbb{R}_{>0}$ and $m\in\mathbb{Z}_{\geq1}$. By
the identifiability property of the $\textrm{Geometric}\left(p\right)$
kernel,\footnote{See Lüxmann-Ellinghaus (1987).} we know that $f_{X\mid c=a}^{\left(\textrm{Q}\right)}\left(m\right)$
is the unique mixing distribution that generates $f_{X\mid a}^{\left(\textrm{ZT}\right)}\left(x\right)$.

One practical benefit of expressing $\textrm{Zeta Tail}\left(a\right)$
as a $\textrm{Geometric}\left(p\right)$ mixture is that it provides
a convenient ``workaround'' for applying the recursion method of
Panjer (1981) to compute CDFs of compound total damage-amount random
variables of the form
\[
T={\displaystyle \sum_{j=1}^{X_{\textrm{ZT}}}}L_{j},
\]
where $X_{\textrm{ZT}}\sim F_{X\mid a}^{\left(\textrm{ZT}\right)}\left(x\right)$
denotes the total number of damage events and the $L_{j}\overset{\textrm{IID}}{\sim}F_{L}\left(\ell\right),\:\ell\in\mathbb{R}_{\geq0}$
(independent of $X_{\textrm{ZT}}$) denote individual damage amounts.
Although the $\textrm{Zeta Tail}\left(a\right)$ PMF fails to satisfy
the basic recurrence
\[
f_{X}\left(x\right)=\left(\alpha+\dfrac{\beta}{x}\right)f_{X}\left(x-1\right),\quad x\in\mathbb{Z}_{\geq2},\alpha+\beta\in\mathbb{R}_{\geq0}
\]
necessary for a direct application of Panjer's algorithm, the $\textrm{Geometric}\left(p\right)$
mixture in (10) can be handled by the more general approach of Tzaninis
and Bozikas (2026).

\section{Applications in Risk Analysis}

\noindent We now consider the $\textrm{Zeta Tail}\left(a\right)$
distribution as an alternative to $\textrm{Geometric}\left(p\right)$
for modeling event counts in risk analysis. Since event-count distributions
often are defined on the sample space $\mathbb{Z}_{\geq0}$ (rather
than $\mathbb{Z}_{\geq1}$) to allow for the possibility of $0$ counts,
we will work with shifted versions of the original formulations; that
is, $\widetilde{X}\mid a\sim\textrm{Zeta Tail 0}\left(a\right)$ with
PMF
\[
f_{\widetilde{X}\mid a}^{\left(\textrm{ZT0}\right)}\left(x\right)=a\zeta\left(x+2,a+1\right),\quad x\in\mathbb{Z}_{\geq0},a\in\mathbb{R}_{>0};
\]
and $\widetilde{X}\mid p\sim\textrm{Geometric 0}\left(p\right)$ with
PMF
\[
f_{\widetilde{X}\mid p}^{\left(\textrm{G0}\right)}\left(x\right)=p\left(1-p\right)^{x},\quad x\in\mathbb{Z}_{\geq0},p\in\left(0,1\right).
\]

The first three subsections below identify specific characteristics
of the $\textrm{Zeta Tail 0}\left(a\right)$ distribution -- its
relative overdispersion (compared to its mean), proportion of $0$
counts (i.e., $\Pr\left\{ \widetilde{X}=0\right\} $), and hazard-function
behavior, respectively -- that clearly differentiate this distribution
from $\textrm{Geometric 0}\left(p\right)$, making it a potentially
better modeling choice in certain situations. Then, in Subsection
4.4, we illustrate the importance of such theoretical differences
by fitting both distributions (along with two others) to a set of
South Korean torrential-rainfall data identified by Lee and Kim (2017)
as having relatively large proportions of $0$ counts.

\subsection{Relative Overdispersion}

\begin{singlespace}
\phantom{}

\medskip{}

\end{singlespace}

\noindent Let
\[
\mathcal{O}_{\widetilde{X}}\left[\widetilde{X}\right]=\dfrac{\textrm{Var}_{\widetilde{X}}\left[\widetilde{X}\right]}{\textrm{E}_{\widetilde{X}}\left[\widetilde{X}\right]}-1
\]
denote the overdispersion index of the random variable $\widetilde{X}\sim F_{\widetilde{X}}\left(x\right),\:x\in\mathbb{Z}_{\geq0}$,
where $\widetilde{X}$ is said to be ``overdispersed'' if $\mathcal{O}_{\widetilde{X}}\left[\widetilde{X}\right]>0$.
This property is important in risk analysis because one often needs
to distinguish between observations following the ubiquitous $\textrm{Poisson}\left(\lambda\right)$
law (for which $\mathcal{O}_{\widetilde{X}}\left[\widetilde{X}\right]=0$)
and those following a $\textrm{Poisson}\left(\lambda\right)$ mixture
(for which $\mathcal{O}_{\widetilde{X}}\left[\widetilde{X}\right]>0$).
Among such mixtures, the most common is $\widetilde{X}\mid r,p=\beta/\left(\beta+1\right)\sim\textrm{Negative Binomial 0}\left(r,p=\beta/\left(\beta+1\right)\right)$,
generated by the continuous mixing distribution $\lambda\mid r,\beta\sim\textrm{Gamma}\left(r,\beta\right)$.

For the special case of $r=1$, $\textrm{Negative Binomial 0}\left(r=1,p=\beta/\left(\beta+1\right)\right)$
is equivalent to\linebreak{}
$\textrm{Geometric 0}\left(p=\beta/\left(\beta+1\right)\right)$,
with
\[
\textrm{E}_{\widetilde{X}\mid p=\tfrac{\beta}{\beta+1}}^{\left(\textrm{G0}\right)}\left[\widetilde{X}\right]=\dfrac{1}{\beta}
\]
and
\[
\textrm{Var}_{\widetilde{X}\mid p=\tfrac{\beta}{\beta+1}}^{\left(\textrm{G0}\right)}\left[\widetilde{X}\right]=\dfrac{\beta+1}{\beta^{2}}.
\]
This implies
\[
\mathcal{O}_{\widetilde{X}\mid p=\tfrac{\beta}{\beta+1}}^{\left(\textrm{G0}\right)}\left[\widetilde{X}\right]=\dfrac{1}{\beta}=\textrm{E}_{\widetilde{X}\mid p=\tfrac{\beta}{\beta+1}}^{\left(\textrm{G0}\right)}\left[\widetilde{X}\right],
\]
and motivates consideration of the distribution's \emph{relative}
overdispersion (compared to its mean):
\[
\dfrac{\mathcal{O}_{\widetilde{X}\mid p=\tfrac{\beta}{\beta+1}}^{\left(\textrm{G0}\right)}\left[\widetilde{X}\right]}{\textrm{E}_{\widetilde{X}\mid p=\tfrac{\beta}{\beta+1}}^{\left(\textrm{G0}\right)}\left[\widetilde{X}\right]}=1.
\]
Adjusting (8) and (9) for the change in sample space (from $\mathbb{Z}_{\geq1}$
to $\mathbb{Z}_{\geq0}$), we find the overdispersion index of $\textrm{Zeta Tail 0}\left(a\right)$
to be
\[
\mathcal{O}_{\widetilde{X}\mid a}^{\left(\textrm{ZT0}\right)}\left[\widetilde{X}\right]=\dfrac{a\left[\zeta\left(2,a\right)+2\zeta\left(3,a\right)-a\left(\zeta\left(2,a\right)\right)^{2}\right]}{a\zeta\left(2,a\right)-1}-1,
\]
with a relative overdispersion of
\begin{equation}
\dfrac{\mathcal{O}_{\widetilde{X}\mid a}^{\left(\textrm{ZT0}\right)}\left[\widetilde{X}\right]}{\textrm{E}_{\widetilde{X}\mid a}^{\left(\textrm{ZT0}\right)}\left[\widetilde{X}\right]}=\dfrac{2a\zeta\left(3,a\right)-a^{2}\left(\zeta\left(2,a\right)\right)^{2}+1}{\left[a\zeta\left(2,a\right)-1\right]^{2}}.
\end{equation}

As shown in Figure 1, the plot of (11) increases from a lower bound
of $1.0$ as $a\rightarrow0^{+}$ to a maximum of about $1.7196$
at $a\approx2.1295$, and then tapers off to $5/3\approx1.6667$ in
the limit as $a\rightarrow\infty$. Thus, the $\textrm{Zeta Tail 0}\left(a\right)$
distribution's level of relative overdispersion is distinctly greater
than that of the $\textrm{Geometric 0}\left(p\right)$ distribution
(which always equals $1.0$) outside a small neighborhood of $0$.
This clear difference provides one simple way of assessing whether
a sample of observed event counts is better modeled by the former
or the latter distribution.

\medskip{}
\medskip{}

\begin{center}
\includegraphics[scale=0.4]{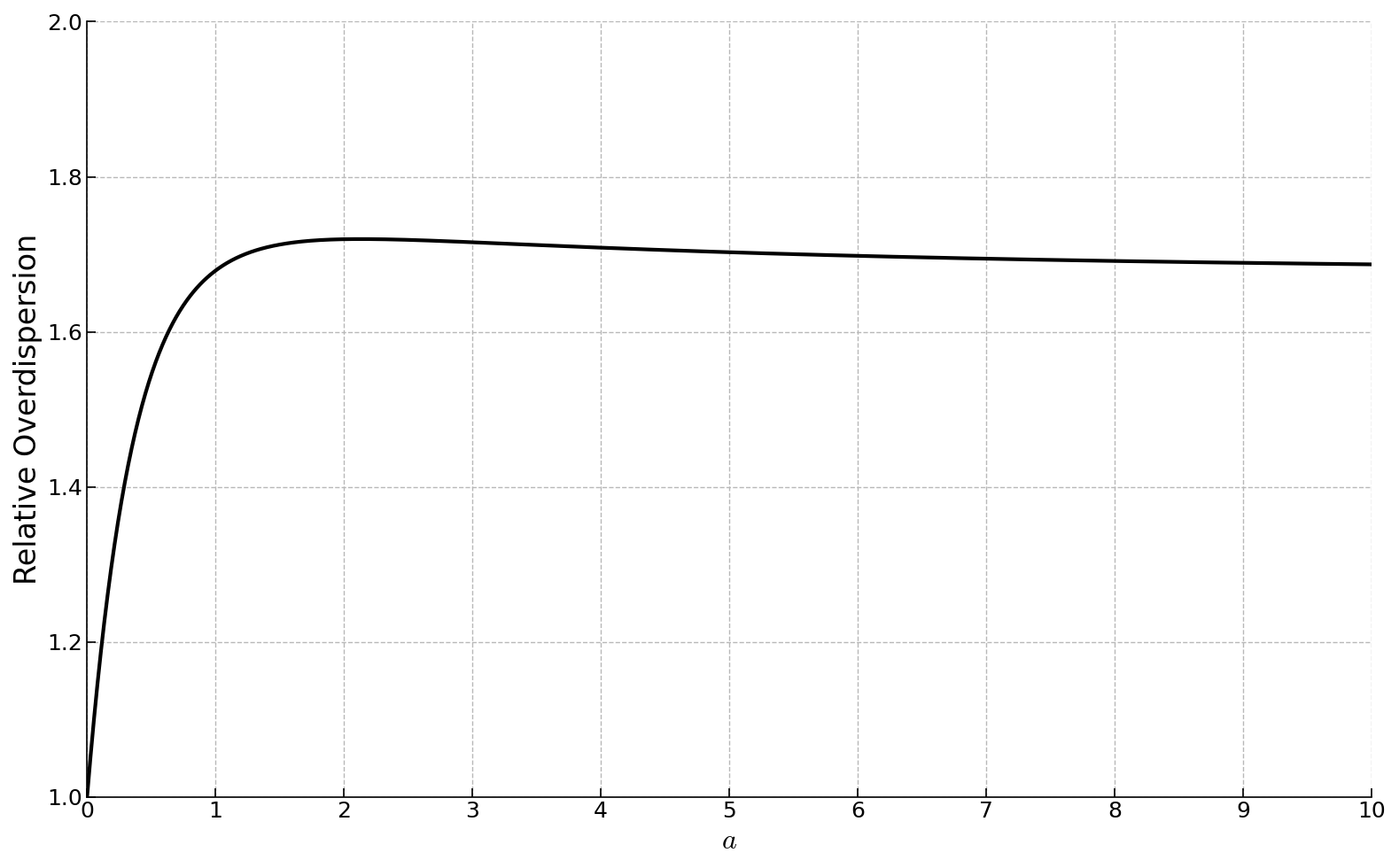}
\par\end{center}

\begin{center}
Figure 1. Relative Overdispersion of $\textrm{Zeta Tail 0}\left(a\right)$\medskip{}
\par\end{center}

\subsection{Proportion of Zero Counts}

\begin{singlespace}
\phantom{}

\medskip{}

\end{singlespace}

\noindent The $\textrm{Zeta Tail 0}\left(a\right)$ distribution's
greater relative overdispersion indicates a more pronounced allocation
of probability to the left and right extremes of the PMF. Thus, it
is not surprising that $\textrm{Zeta Tail 0}\left(a\right)$ assigns
relatively more probability to the outcome $x=0$ than $\textrm{Geometric 0}\left(p\right)$.
To see this, one can fix the two distributions' parameter values so
they both possess the same tail behavior (i.e., set $p=a/\left(a+1\right)$),
and then compute the ratio of their probability masses at $x=0$:
\[
\dfrac{f_{\widetilde{X}\mid a}^{\left(\textrm{ZT0}\right)}\left(0\right)}{f_{\widetilde{X}\mid p=\tfrac{a}{a+1}}^{\left(\textrm{G0}\right)}\left(0\right)}=\dfrac{a\zeta\left(2,a+1\right)}{\left(\dfrac{a}{a+1}\right)}
\]
\[
=\left(a+1\right)\zeta\left(2,a+1\right).
\]

\medskip{}
\medskip{}

\begin{center}
\includegraphics[scale=0.4]{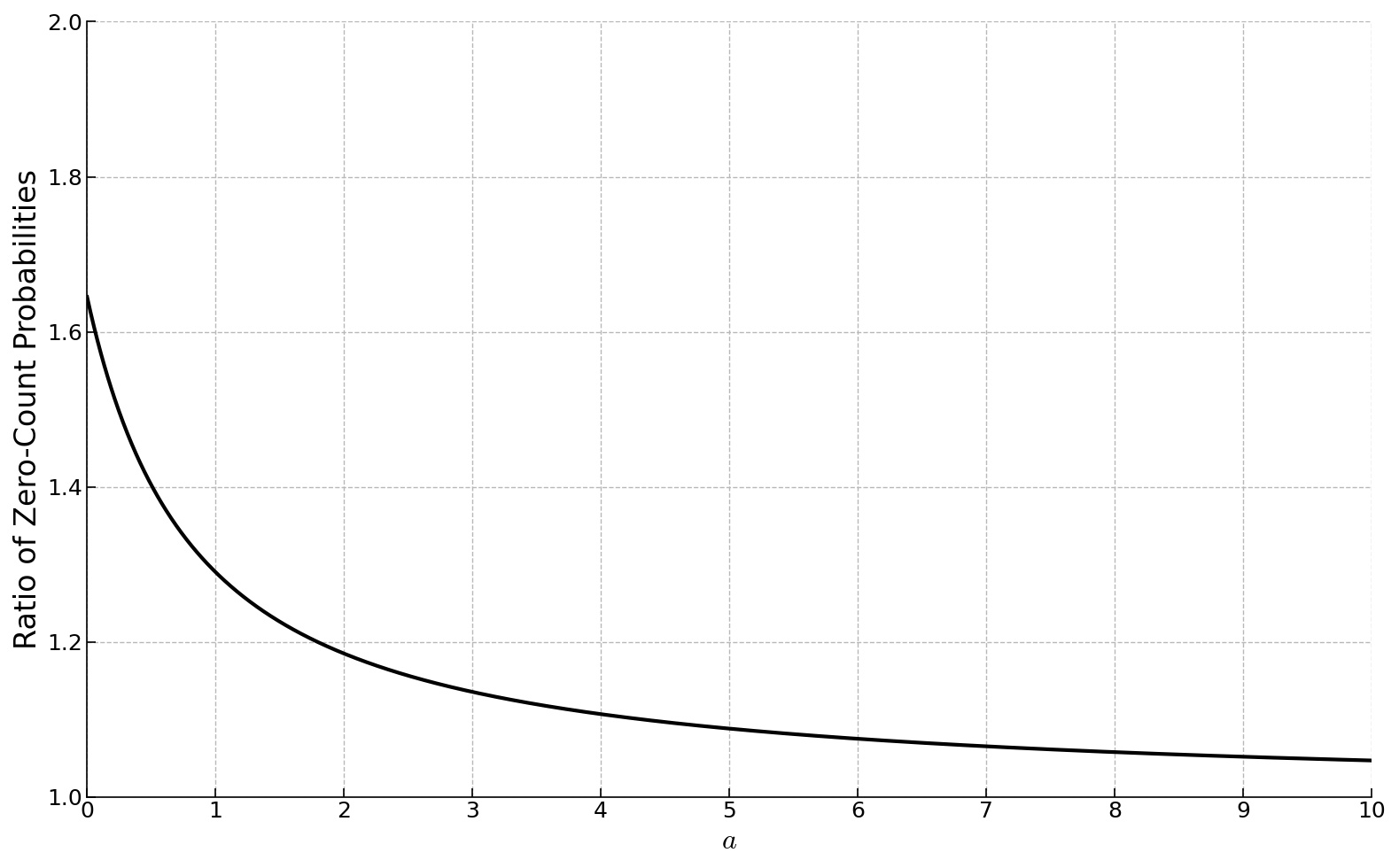}
\par\end{center}

\begin{center}
Figure 2. Ratio of Zero-Count Probabilities ($\textrm{Zeta Tail 0}\left(a\right)$
vs. $\textrm{Geometric 0}\left(p=\tfrac{a}{a+1}\right)$)\medskip{}
\par\end{center}

This ratio, plotted in Figure 2, decreases in $a$ from an upper bound
of $\zeta\left(2\right)\approx1.6449$ as $a\rightarrow0^{+}$ to
a limiting value of $1.0$ as $a\rightarrow\infty$. In other words,
for small values of $a$, the $\textrm{Zeta Tail 0}\left(a\right)$
distribution raises the probability of $0$ counts over that of the
$\textrm{Geometric 0}\left(p\right)$ distribution -- assuming comparable
tail characteristics -- by almost $65$ percent. In the actuarial
finance literature, zero-inflated event-count distributions are particularly
useful for modeling insurance loss frequencies, which often arise
from implicit mixtures of two random components: (i) a fundamental
process generating the actual number of damages occurring within a
given time period; and (ii) a secondary process causing certain damage
occurrences to be removed from the first process (e.g., through enhanced
risk-control measures or the application of insurance deductibles
to preclude payment). The $\textrm{Zeta Tail 0}\left(a\right)$ distribution's
naturally greater zero-count probability could make it more attractive
in such settings.

\subsection{Hazard-Function Behavior}

\begin{singlespace}
\phantom{}

\medskip{}

\end{singlespace}

\noindent In Proposition 3, we found that the $\textrm{Zeta Tail}\left(a\right)$
distribution's hazard function is strictly decreasing in $x$, tending
to a limit of $a$ as $x\rightarrow\infty$. These characteristics
remain unchanged for $\textrm{Zeta Tail 0}\left(a\right)$, and are
illustrated in Figure 3 for $a=0.25$, $1.0$, and $4.0$. On the
other hand, the $\textrm{Geometric 0}\left(p\right)$ hazard function
is constant over $x$ as a consequence of that distribution's memoryless
property; specifically,
\[
h_{\widetilde{X}\mid p}^{\left(\textrm{G0}\right)}\left(x\right)=\dfrac{f_{\widetilde{X}\mid p}^{\left(\textrm{G0}\right)}\left(x\right)}{1-F_{\widetilde{X}\mid p}^{\left(\textrm{G0}\right)}\left(x\right)}
\]
\[
=\dfrac{p\left(1-p\right)^{x}}{{\displaystyle \sum_{k=x+1}^{\infty}}p\left(1-p\right)^{k}}
\]
\[
=\dfrac{p}{1-p}.
\]

\medskip{}
\noindent\medskip{}
\noindent{}
\begin{center}
\includegraphics[scale=0.4]{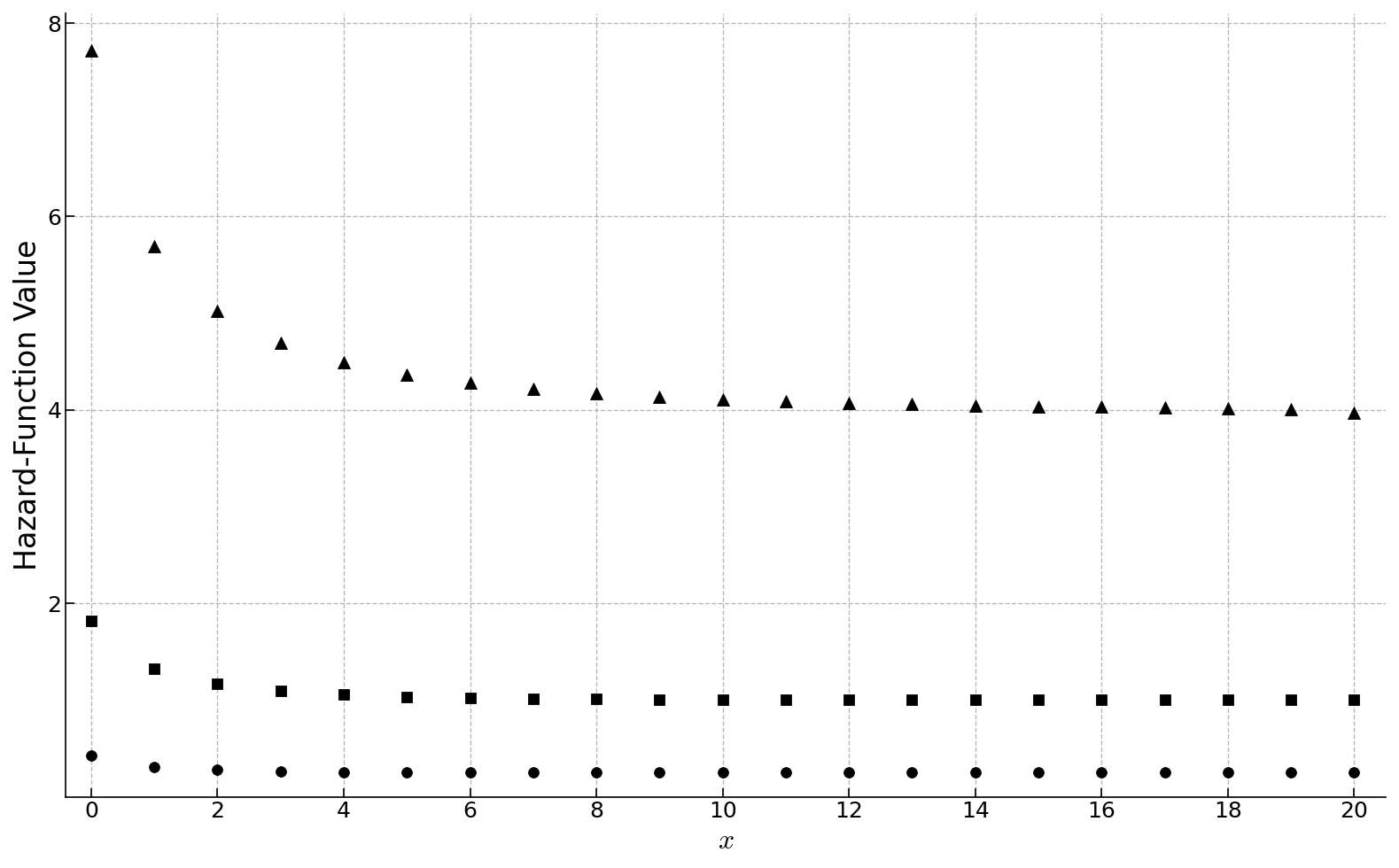}
\par\end{center}

\begin{center}
Figure 3. Hazard Functions of $\textrm{Zeta Tail 0}\left(a\right)$
\par\end{center}

\begin{center}
(circles $\Longrightarrow a=0.25$; squares $\Longrightarrow a=1.0$;
triangles $\Longrightarrow a=4.0$)\medskip{}
\par\end{center}

This qualitative difference affords an additional means of selecting
between the two distributions for an appropriate model of observed
event-count data. Essentially, for a given sample of IID event counts,
$\widetilde{X}_{j},\:j\in\left\{ 1,2,\ldots,N\right\} $, one can
compute the empirical hazard function,
\[
\widehat{h}_{N}\left(x\right)=\dfrac{\textrm{Number of }j\textrm{ such that }\widetilde{X}_{j}=x}{\textrm{Number of }j\textrm{ such that }\widetilde{X}_{j}>x},\quad x\in\left\{ 0,1,\ldots,\underset{j}{\max}\left\{ \widetilde{X}_{j}\right\} -1\right\} ,
\]
and determine which of $h_{\widetilde{X}\mid a}^{\left(\textrm{ZT0}\right)}\left(x\right)$
or $h_{\widetilde{X}\mid p}^{\left(\textrm{G0}\right)}\left(x\right)$
it more closely resembles. If $\widehat{h}_{N}\left(x\right)$ tends
to decrease in $x$, then adopting $\textrm{Zeta Tail 0}\left(a\right)$
offers a simpler alternative to adding a second parameter to the $\textrm{Geometric 0}\left(p\right)$
model (e.g., by employing $\textrm{Negative Binomial 0}\left(r,p\right)$
for $r\in\left(0,1\right)$\footnote{The hazard function of $\textrm{Negative Binomial 0}\left(r,p\right)$
is strictly decreasing for $r\in\left(0,1\right)$ and strictly increasing
for $r\in\left(1,\infty\right)$.}).

Given that the empirical mean residual-life function,
\[
\widehat{e}_{N}\left(x\right)=\dfrac{{\displaystyle \sum_{j:\widetilde{X}_{j}>x}}\left(\widetilde{X}_{j}-x\right)}{\textrm{Number of }j\textrm{ such that }\widetilde{X}_{j}>x},\quad x\in\left\{ 0,1,\ldots,\underset{j}{\max}\left\{ \widetilde{X}_{j}\right\} -1\right\} ,
\]
embodies information similar to that encoded in $\widehat{h}_{N}\left(x\right)$,
the former measure offers a comparable means of differentiating between
alternative modeling distributions. In this case, if $\widehat{e}_{N}\left(x\right)$
tends to increase in $x$, then adopting $\textrm{Zeta Tail 0}\left(a\right)$
offers a simple 1-parameter alternative to the $\textrm{Geometric 0}\left(p\right)$
model.

\subsection{Torrential-Rainfall Counts}

\begin{singlespace}
\phantom{}

\medskip{}

\end{singlespace}

\noindent The preceding three subsections identified specific characteristics
of the $\textrm{Zeta Tail 0}\left(a\right)$ distribution that make
it a potentially superior alternative to $\textrm{Geometric 0}\left(p\right)$
for certain types of data. To illustrate this point, we now consider
a set of meteorological event-count data distinguished by relatively
large proportions of $0$ counts, and compare goodness-of-fit criteria
computed from fitting four distinct distributions: $\textrm{Geometric 0}\left(p\right)$,
$\textrm{Zero-Inflated Geometric 0}\left(p,\pi_{0}\right)$, $\textrm{Negative Binomial 0}\left(r,p\right)$,
and $\textrm{Zeta Tail 0}\left(a\right)$, where:

\begin{singlespace}
$\bullet$ the $\textrm{Zero-Inflated Geometric 0}\left(p,\pi_{0}\right)$
distribution is characterized by PMF
\[
f_{\widetilde{X}\mid p,\pi_{0}}^{\left(\textrm{ZIG0}\right)}\left(x\right)=\begin{cases}
\pi_{0}+\left(1-\pi_{0}\right)p, & x=0,p\in\left(0,1\right),\pi_{0}\in\left[0,1\right)\\
\left(1-\pi_{0}\right)p\left(1-p\right)^{x}, & x\in\mathbb{Z}_{\geq1},p\in\left(0,1\right),\pi_{0}\in\left[0,1\right)
\end{cases},
\]
with $\pi_{0}$ denoting the probability of observing a fixed event
count of $0$, and $1-\pi_{0}$ denoting the probability of observing
a conventional $\textrm{Geometric 0}\left(p\right)$ random variable;
and

$\bullet$ the $\textrm{Negative Binomial 0}\left(r,p\right)$ distribution
is characterized by PMF
\[
f_{\widetilde{X}\mid r,p}^{\left(\textrm{NB0}\right)}\left(x\right)=\dfrac{\Gamma\left(r+x\right)}{\Gamma\left(r\right)\Gamma\left(x+1\right)}p^{r}\left(1-p\right)^{x},\quad x\in\mathbb{Z}_{\geq0},r\in\mathbb{R}_{>0},p\in\left(0,1\right).
\]

\end{singlespace}

The data, published in a study by Lee and Kim (2017), consist of monthly
torrential-rainfall counts for each of $12$ distinct regions in South
Korea compiled over a period of $32$ years (1983-2014). Since the
primary objective of Lee and Kim (2017) -- to construct a Bayesian
estimation model based on Poisson and modified-Poisson sampling distributions
-- was quite different from ours, we will not make direct comparisons
between the respective analyses except to note that Geometric-based
sampling distributions appear to fit the raw data better, on the whole,
than Poisson-based distributions.

Table 2 summarizes the monthly torrential-rainfall counts, along with
annual mean rainfall measurements that can be used to divide the $12$
regions into two clusters: six ``drier'' regions consisting of rows
(1), (3), (4), (5), (6), and (7) and six ``wetter'' regions consisting
of rows (2), (8), (9), (10), (11), and (12). For data-analytic purposes,
it also is important to note that the maximum number of counts from
any region within all of the $384$ monthly periods is exactly $6$
(as shown), so we need not worry about truncation or censoring artifacts
in fitting probability models.\medskip{}

\begin{center}
Table 2. Torrential-Rainfall Event Counts from $12$ South Korean
Regions
\par\end{center}

\begin{center}
\begin{tabular}{|c|c|c|c|c|c|c|c|c|c|}
\hline 
{\small\textbf{Region}} & {\small\textbf{Annual Mean}} & \multicolumn{8}{c|}{{\small\textbf{Monthly Event Counts (1983-2014)}}}\tabularnewline
\cline{3-10}
 & {\small\textbf{Rainfall (mm)}} & {\small\textbf{\quad{}$\boldsymbol{0}$\quad{}}} & {\small\textbf{\quad{}$\boldsymbol{1}$\quad{}}} & {\small\textbf{\quad{}$\boldsymbol{2}$\quad{}}} & {\small\textbf{\quad{}$\boldsymbol{3}$\quad{}}} & {\small\textbf{\quad{}$\boldsymbol{4}$\quad{}}} & {\small\textbf{\quad{}$\boldsymbol{5}$\quad{}}} & {\small\textbf{\quad{}$\boldsymbol{6}$\quad{}}} & {\small\textbf{\# Months}}\tabularnewline
\hline 
{\small\textbf{(1) Daegu}} & {\small$1087$} & {\small$309$} & {\small$53$} & {\small$18$} & {\small$2$} & {\small$2$} & {\small$0$} & {\small$0$} & {\small$384$}\tabularnewline
\hline 
{\small\textbf{(2) Busan}} & {\small$1344$} & {\small$275$} & {\small$67$} & {\small$32$} & {\small$5$} & {\small$4$} & {\small$0$} & {\small$1$} & {\small$384$}\tabularnewline
\hline 
{\small\textbf{(3) Yeongju}} & {\small$1125$} & {\small$295$} & {\small$60$} & {\small$25$} & {\small$1$} & {\small$1$} & {\small$1$} & {\small$1$} & {\small$384$}\tabularnewline
\hline 
{\small\textbf{(4) Mungyeong}} & {\small$1095$} & {\small$297$} & {\small$57$} & {\small$22$} & {\small$3$} & {\small$2$} & {\small$2$} & {\small$1$} & {\small$384$}\tabularnewline
\hline 
{\small\textbf{(5) Uiseong}} & {\small$1039$} & {\small$324$} & {\small$42$} & {\small$14$} & {\small$3$} & {\small$1$} & {\small$0$} & {\small$0$} & {\small$384$}\tabularnewline
\hline 
{\small\textbf{(6) Gumi}} & {\small$1104$} & {\small$311$} & {\small$50$} & {\small$15$} & {\small$4$} & {\small$2$} & {\small$1$} & {\small$1$} & {\small$384$}\tabularnewline
\hline 
{\small\textbf{\enskip{}(7) Yeongcheon\enskip{}}} & {\small$1074$} & {\small$328$} & {\small$49$} & {\small$6$} & {\small$1$} & {\small$0$} & {\small$0$} & {\small$0$} & {\small$384$}\tabularnewline
\hline 
{\small\textbf{(8) Geochang}} & {\small$1334$} & {\small$255$} & {\small$80$} & {\small$35$} & {\small$8$} & {\small$4$} & {\small$1$} & {\small$1$} & {\small$384$}\tabularnewline
\hline 
{\small\textbf{(9) Hapcheon}} & {\small$1317$} & {\small$260$} & {\small$87$} & {\small$32$} & {\small$4$} & {\small$1$} & {\small$0$} & {\small$0$} & {\small$384$}\tabularnewline
\hline 
{\small\textbf{(10) Miryang}} & {\small$1249$} & {\small$267$} & {\small$81$} & {\small$27$} & {\small$4$} & {\small$3$} & {\small$1$} & {\small$1$} & {\small$384$}\tabularnewline
\hline 
{\small\textbf{(11) Pohang}} & {\small$1274$} & {\small$281$} & {\small$69$} & {\small$29$} & {\small$2$} & {\small$2$} & {\small$0$} & {\small$1$} & {\small$384$}\tabularnewline
\hline 
{\small\textbf{(12) Ulsan}} & {\small$1287$} & {\small$277$} & {\small$78$} & {\small$24$} & {\small$2$} & {\small$2$} & {\small$1$} & {\small$0$} & {\small$384$}\tabularnewline
\hline 
\end{tabular}\medskip{}
\medskip{}
\par\end{center}

In Table 3, we present two goodness-of-fit criteria for each of the
four indicated distributions for each region: (i) the $\textrm{Pearson's }\chi^{2}\textrm{ }p\textrm{-value}$,
for which the associated $\chi^{2}$ statistic is computed using $4$
bins (for $0$, $1$, $2$, and $3$+ counts, respectively) and all
parameters are estimated by maximum likelihood; and (ii) the Akaike
information criterion (AIC). Then, for each region, the highest $\textrm{Pearson's }\chi^{2}\:p\textrm{-value}$
and the lowest AIC value are marked by asterisks.
\begin{center}
\newpage{}
\par\end{center}

\begin{center}
Table 3. Torrential-Rainfall Event-Count Model Comparisons
\par\end{center}

\begin{center}
\begin{tabular}{|c|c|c|c|c|c|c|c|c|}
\hline 
{\small\textbf{Region}} & \multicolumn{8}{c|}{{\small\textbf{Competing Models}}}\tabularnewline
\cline{2-9}
 & \multicolumn{2}{c|}{{\small\textbf{$\boldsymbol{\textrm{Geom. 0}\left(p\right)}$}}} & \multicolumn{2}{c|}{{\small\textbf{$\boldsymbol{\textrm{ZI Geom. 0}\left(p,\pi_{0}\right)}$}}} & \multicolumn{2}{c|}{{\small\textbf{$\boldsymbol{\textrm{Neg. Bin. 0}\left(r,p\right)}$}}} & \multicolumn{2}{c|}{{\small\textbf{$\boldsymbol{\textrm{Zeta Tail 0}\left(a\right)}$}}}\tabularnewline
\cline{2-9}
 & {\small\textbf{$\boldsymbol{\chi_{2}^{2}\;\:p\textrm{-val.}}$}} & {\small\textbf{AIC}} & {\small\textbf{$\boldsymbol{\chi_{1}^{2}\;\:p\textrm{-val.}}$}} & {\small\textbf{AIC}} & {\small\textbf{$\boldsymbol{\chi_{1}^{2}\;\:p\textrm{-val.}}$}} & {\small\textbf{AIC}} & {\small\textbf{$\boldsymbol{\chi_{2}^{2}\;\:p\textrm{-val.}}$}} & {\small\textbf{AIC}}\tabularnewline
\hline 
{\small\textbf{(1) Daegu}} & {\small\textbf{$0.1709$}} & {\small\textbf{$504.52$}} & {\small\textbf{$0.2842$}} & {\small\textbf{$503.79$}} & {\small\textbf{$0.2021$}} & {\small\textbf{$504.21$}} & {\small\textbf{$0.4028^{*}$}} & {\small\textbf{$502.38^{*}$}}\tabularnewline
\hline 
{\small\textbf{(2) Busan}} & {\small\textbf{$0.0847$}} & {\small\textbf{$680.41$}} & {\small\textbf{$0.0762$}} & {\small\textbf{$679.94$}} & {\small\textbf{$0.0475$}} & {\small\textbf{$680.57$}} & {\small\textbf{$0.1003^{*}$}} & {\small\textbf{$679.10^{*}$}}\tabularnewline
\hline 
{\small\textbf{(3) Yeongju}} & {\small\textbf{$0.0621$}} & {\small\textbf{$577.83$}} & {\small\textbf{$0.0390$}} & {\small\textbf{$577.19$}} & {\small\textbf{$0.0243$}} & {\small\textbf{$577.30$}} & {\small\textbf{$0.0683^{*}$}} & {\small\textbf{$575.34^{*}$}}\tabularnewline
\hline 
{\small\textbf{(4) Mungyeong}} & {\small\textbf{$0.0810$}} & {\small\textbf{$594.26$}} & {\small\textbf{$0.3351$}} & {\small\textbf{$588.59$}} & {\small\textbf{$0.2291$}} & {\small\textbf{$588.23$}} & {\small\textbf{$0.4348^{*}$}} & {\small\textbf{$587.43^{*}$}}\tabularnewline
\hline 
{\small\textbf{(5) Uiseong}} & {\small\textbf{$0.0307$}} & {\small\textbf{$439.05$}} & {\small\textbf{$0.5023^{*}$}} & {\small\textbf{$434.82^{*}$}} & {\small\textbf{$0.3250$}} & {\small\textbf{$435.49$}} & {\small\textbf{$0.2524$}} & {\small\textbf{$435.06$}}\tabularnewline
\hline 
{\small\textbf{(6) Gumi}} & {\small\textbf{$0.0191$}} & {\small\textbf{$528.90$}} & {\small\textbf{$0.6213$}} & {\small\textbf{$520.19$}} & {\small\textbf{$0.7797^{*}$}} & {\small\textbf{$519.43^{*}$}} & {\small\textbf{$0.3551$}} & {\small\textbf{$520.71$}}\tabularnewline
\hline 
{\small\textbf{\enskip{}(7) Yeongcheon\enskip{}}} & {\small\textbf{$0.9155^{*}$}} & {\small\textbf{$369.46^{*}$}} & {\small\textbf{$0.6743$}} & {\small\textbf{$371.46$}} & {\small\textbf{$0.8445$}} & {\small\textbf{$371.26$}} & {\small\textbf{$0.4783$}} & {\small\textbf{$370.98$}}\tabularnewline
\hline 
{\small\textbf{(8) Geochang}} & {\small\textbf{$0.4446^{*}$}} & {\small\textbf{$754.76^{*}$}} & {\small\textbf{$0.2285$}} & {\small\textbf{$756.48$}} & {\small\textbf{$0.2057$}} & {\small\textbf{$756.61$}} & {\small\textbf{$0.1055$}} & {\small\textbf{$757.67$}}\tabularnewline
\hline 
{\small\textbf{(9) Hapcheon}} & {\small\textbf{$0.0520$}} & {\small\textbf{$678.03$}} & {\small\textbf{$0.0150$}} & {\small\textbf{$680.03$}} & {\small\textbf{$0.0679^{*}$}} & {\small\textbf{$676.27^{*}$}} & {\small\textbf{$0.0008$}} & {\small\textbf{$687.99$}}\tabularnewline
\hline 
{\small\textbf{(10) Miryang}} & {\small\textbf{$0.7582^{*}$}} & {\small\textbf{$685.15^{*}$}} & {\small\textbf{$0.4234$}} & {\small\textbf{$687.13$}} & {\small\textbf{$0.3837$}} & {\small\textbf{$687.08$}} & {\small\textbf{$0.1102$}} & {\small\textbf{$687.65$}}\tabularnewline
\hline 
{\small\textbf{(11) Pohang}} & {\small\textbf{$0.0880^{*}$}} & {\small\textbf{$628.52^{*}$}} & {\small\textbf{$0.0318$}} & {\small\textbf{$630.01$}} & {\small\textbf{$0.0260$}} & {\small\textbf{$630.19$}} & {\small\textbf{$0.0315$}} & {\small\textbf{$629.61$}}\tabularnewline
\hline 
{\small\textbf{(12) Ulsan}} & {\small\textbf{$0.4584^{*}$}} & {\small\textbf{$623.36^{*}$}} & {\small\textbf{$0.2117$}} & {\small\textbf{$625.36$}} & {\small\textbf{$0.2770$}} & {\small\textbf{$625.21$}} & {\small\textbf{$0.0485$}} & {\small\textbf{$627.46$}}\tabularnewline
\hline 
\end{tabular}
\par\end{center}

\begin{center}
{\small$^{*}$ Denotes the highest $\chi^{2}\:p\textrm{-value}$ or
lowest AIC value within a given region.}{\small\par}
\par\end{center}

\medskip{}

On the whole, the goodness-of-fit results in Table 3 indicate that
the two $1$-parameter models ($\textrm{Geometric 0}\left(p\right)$
and $\textrm{Zeta Tail 0}\left(a\right)$) perform slightly better
than their $2$-parameter competitors. In fact, under both goodness-of-fit
criteria, the $\textrm{Geometric 0}\left(p\right)$ distribution provides
the best fit in $5$ of the $12$ regions and the $\textrm{Zeta Tail 0}\left(a\right)$
distribution does best in $4$ of the $12$ regions. Moreover, $\textrm{Geometric 0}\left(p\right)$
appears the most robust of all four models, with only $2$ significant
$\textrm{Pearson's }\chi^{2}$ statistics beyond the $0.05$ level.
Examining the goodness-of-fit measures more closely, one can see further
that $\textrm{Geometric 0}\left(p\right)$ tends to do relatively
better within the cluster of ``wetter'' regions, where the proportions
of $0$ counts are smaller and $\textrm{Zeta Tail 0}\left(a\right)$
enjoys no special advantage.

\section{Further Modeling Considerations}

\noindent The theoretical analyses and illustrative application of
the previous section demonstrate clearly that, given event-count data
with certain known characteristics, the $\textrm{Zeta Tail 0}\left(a\right)$
distribution can provide a better model than both $\textrm{Geometric 0}\left(p\right)$
and certain $2$-parameter alternatives. However, it also is important
to consider how useful $\textrm{Zeta Tail 0}\left(a\right)$ is generally,
without conditioning on the particular characteristics of an observed
sample. To this end, we apply the formal measure proposed by Powers
and Xu (2025a) to quantify the \emph{versatility} of a potential distributional
model, where versatility is defined (conceptually) as a combination
of two desirable characteristics of a probability distribution: \emph{simplicity}
-- the degree of plainness or coherence displayed by the probability
distribution's mathematical form (which is identified as the opposite
of \emph{complexity}); and \emph{adaptability} -- the degree of naturalness
or resiliency inherent in the probability distribution's mathematical
form (identified as the opposite of \emph{contrivance}).

For a given (discrete or continuous) probability distribution $F_{X\mid\boldsymbol{\theta}}\left(x\right)$
with $k$-dimensional parameter vector $\boldsymbol{\theta}=\left[\theta_{1},\theta_{2},\ldots,\theta_{k}\right]\in\left(\mathbb{R}_{>0}\right)^{k}$,
Powers and Xu (2025a) argued that the determinant of the Fisher information
matrix is positively related to the versatility of the distribution.
To account for all possible values of $\boldsymbol{\theta}$, they
worked with the Bayesian Fisher information matrix (BFIM)\footnote{See, e.g., Daniels and Hogan (2008).}
computed by averaging over a maximum-entropy joint prior distribution
of the $k$ parameters (specifically, $\theta_{1},\theta_{2},\ldots,\theta_{k}\overset{\textrm{IID}}{\sim}\textrm{Lognormal}\left(\mu=0,\sigma=1\right)$).
Normalizing for the number of parameters by raising the determinant
of the BFIM to the $1/\left(2k\right)$ power, they thus proposed
the versatility measure
\[
\mathcal{V}\left(F_{X\mid\boldsymbol{\theta}}\right)=\left(\det\left(\textrm{E}_{P\left(\boldsymbol{\theta}\right)}\left[\mathbf{I}_{F_{X\mid\boldsymbol{\theta}}}^{\left(k\right)}\left(\boldsymbol{\theta}\right)\right]\right)\right)^{1/\left(2k\right)},
\]
where $\mathbf{I}_{F_{X\mid\boldsymbol{\theta}}}^{\left(k\right)}\left(\boldsymbol{\theta}\right)$
is the Fisher information matrix and $\textrm{E}_{P\left(\boldsymbol{\theta}\right)}\left[\cdot\right]$
denotes the expected-value operator of the joint prior distribution.
For the $1$-parameter case, this simplifies to
\begin{equation}
\mathcal{V}\left(F_{X\mid\theta}\right)=\sqrt{{\displaystyle \int_{0}^{\infty}}\textrm{E}_{X\mid\theta}\left[\left(\dfrac{\partial\ln\left(f_{X\mid\theta}\left(X\right)\right)}{\partial\theta}\right)^{2}\right]\dfrac{1}{\sqrt{2\pi}\theta}\exp\left(-\dfrac{\left(\ln\left(\theta\right)\right)^{2}}{2}\right)d\theta}.
\end{equation}

Table 4 presents computations of (12) for the $\textrm{Geometric 0}\left(p\left(m\right)\right),\:m\in\mathbb{R}_{>0}$
and $\textrm{Zeta Tail 0}\left(a\right)$ probability distributions
consistent with Powers and Xu (2025a). However, we will not use the
content of the shaded cells for the reasons that follow.\medskip{}

\begin{center}
Table 4. $\textrm{Geometric 0}\left(p\left(m\right)\right)$ and $\textrm{Zeta Tail 0}\left(a\right)$
Versatility Measures ($\mathcal{V}$)$^{*}$
\par\end{center}

\begin{center}
\begin{tabular}{|c|c|c|}
\hline 
$\textrm{Geometric 0}\left(p_{1}\left(m\right)=\tfrac{m}{m+1}\right)$ & \cellcolor{gray!25}$\textrm{Geometric 0}\left(p_{2}\left(m\right)=\tfrac{1}{m+1}\right)$ & $\quad\quad\quad\textrm{Zeta Tail 0}\left(a\right)\quad\quad\quad$\tabularnewline
\hline 
$\mathcal{V}=2.4981$ & \cellcolor{gray!25}$\mathcal{V}=1.0718$ & $\mathcal{V}=2.4827$\tabularnewline
\hline 
\multicolumn{2}{|c|}{\cellcolor{gray!25}$\textrm{(Avg. }\mathcal{V}=1.7850\textrm{)}$} & \tabularnewline
\hline 
\end{tabular}
\par\end{center}

\begin{center}
{\small$^{*}$ All versatility measures are consistent with computations
of Powers and Xu (2025a). However, the values in shaded cells are
not used in the present analysis.}\medskip{}
\medskip{}
\par\end{center}

In the $\textrm{Geometric 0}\left(p\left(m\right)\right)$ case, Powers
and Xu (2025a) constructed $\mathcal{V}\left(F_{\widetilde{X}\mid p\left(m\right)}^{\left(\textrm{G0}\right)}\right)$
as the average of two distinct calculations because they noted that
the PMF enjoys two symmetric parameterizations for $m$ -- $p_{1}\left(m\right)=m/\left(m+1\right)$
and $p_{2}\left(m\right)=1/\left(m+1\right)$ -- whose associated
PMFs,
\begin{equation}
f_{\widetilde{X}\mid p_{1}\left(m\right)}^{\left(\textrm{G0}\right)}\left(x\right)=\left(\dfrac{m}{m+1}\right)\left(\dfrac{1}{m+1}\right)^{x},\quad x\in\mathbb{Z}_{\geq0},m\in\mathbb{R}_{>0}
\end{equation}
and
\begin{equation}
f_{\widetilde{X}\mid p_{2}\left(m\right)}^{\left(\textrm{G0}\right)}\left(x\right)=\left(\dfrac{1}{m+1}\right)\left(\dfrac{m}{m+1}\right)^{x},\quad x\in\mathbb{Z}_{\geq0},m\in\mathbb{R}_{>0},
\end{equation}
respectively, appear to be ``equally compressible''. That is, the
expressions in (13) and (14) appear to require the same number of
distinct mathematical symbols to write out in full.

However, employing the symbol-counting system proposed by Powers and
Xu (2025a) themselves, one finds that the ``most efficient'' expressions
for the two alternative PMFs are clearly different. Specifically,
\[
f_{\widetilde{X}\mid p_{1}\left(m\right)}^{\left(\textrm{G0}\right)}\left(x\right)=m/\left(m+1\right)\overset{\wedge}{}\left(x+1\right)
\]
(where ``$\overset{\wedge}{}$'' denotes ``raised to the power
of'') requires $13$ symbols, whereas
\[
f_{\widetilde{X}\mid p_{2}\left(m\right)}^{\left(\textrm{G0}\right)}\left(x\right)=m\overset{\wedge}{}x/\left(m+1\right)\overset{\wedge}{}\left(x+1\right)
\]
requires $15$ symbols. Consequently, according to the authors' formal
procedure, we may ignore the $p_{2}\left(m\right)$ case and simply
compare $\mathcal{V}\left(F_{\widetilde{X}\mid p_{1}\left(m\right)}^{\left(\textrm{G0}\right)}\right)=2.4981$
to $\mathcal{V}\left(F_{\widetilde{X}\mid a}^{\left(\textrm{ZT0}\right)}\right)=2.4827$,
thus finding that the two versatility measures are approximately the
same. Although Powers and Xu (2025a) did not propose a range of ``acceptable''
versatility measures, it is instructive to note that $\mathcal{V}\left(F_{\widetilde{X}\mid a}^{\left(\textrm{ZT0}\right)}\right)=2.4827$
falls near the high end of values for discrete distributions computed
in their article. Consequently, we would conclude that the $\textrm{Zeta Tail 0}\left(a\right)$
PMF constitutes a reasonable alternative to $\textrm{Geometric 0}\left(p\right)$
as a general model of event-count data.

\section{Conclusion}

\noindent In the present study, we introduced the $\textrm{Zeta Tail}\left(a\right)$
probability distribution as a novel model for damage-event counts
in risk analysis. Although not significantly addressed in the scholarly
literature, this distribution is actually quite tractable, permitting
the derivation of closed-form expressions for its moments, generating
functions, and reliability functions. Further, we showed that it admits
a straightforward interpretation as a countable (and necessarily identifiable)
mixture of $\textrm{Geometric}\left(p\right)$ distributions.

By comparing certain properties of the $\textrm{Zeta Tail 0}\left(a\right)$
and $\textrm{Geometric 0}\left(p\right)$ distributions -- including
relative overdispersion (compared to the mean), proportion of $0$
counts, and hazard-function behavior -- we found that the former
model may be preferable to the latter for certain types of event-count
data. Moreover, a broader assessment of the two distributions' overall
applicability (using the versatility measure of Powers and Xu, 2025a)
revealed that $\textrm{Zeta Tail 0}\left(a\right)$ is a reasonable
alternative to $\textrm{Geometric 0}\left(p\right)$ in more general
risk-analytic settings.

\medskip{}
\medskip{}

\medskip{}
\medskip{}

\section*{Appendix}

\begin{singlespace}
\textbf{\medskip{}
}

\noindent\textbf{Proposition A.1.} \emph{The PMF of the $\textrm{Quadratic}\left(c=1\right)$
distribution cannot be written as either a discrete or continuous
mixture of $\textrm{Zeta}\left(b\right)$ PMFs.}
\end{singlespace}
\begin{proof}
\begin{singlespace}
\phantom{}
\end{singlespace}

\begin{singlespace}
\medskip{}

\noindent We first consider the possibility of continuous mixtures,
assuming (for purposes of contradiction) that $f_{X\mid c=1}^{\left(\textrm{Q}\right)}\left(x\right),\:x\in\mathbb{Z}_{\geq1}$
can be written as ${\textstyle \int_{0}^{\infty}}f_{X\mid b}^{\left(\textrm{Z}\right)}\left(x\right)g\left(b\right)db$
for some mixing PDF $g\left(b\right)$, for $b\in\mathbb{R}_{>0}$.
Then
\[
f_{X\mid c=1}^{\left(\textrm{Q}\right)}\left(x\right)=\dfrac{1}{x\left(x+1\right)}={\displaystyle \int_{0}^{\infty}}\dfrac{g\left(b\right)}{\zeta\left(b+1\right)x^{b+1}}db
\]
\[
\Longleftrightarrow\dfrac{2}{x+1}={\displaystyle \int_{0}^{\infty}}\dfrac{2g\left(b\right)}{\zeta\left(b+1\right)x^{b}}db,\qquad\qquad\qquad\textrm{(A.1)}
\]
and we can set $x=1$ in (A.1) to obtain
\[
1={\displaystyle \int_{0}^{\infty}}\dfrac{2g\left(b\right)}{\zeta\left(b+1\right)}db={\displaystyle \int_{0}^{\infty}}\eta\left(b\right)db,\qquad\qquad\qquad\textrm{(A.2)}
\]
where $\eta\left(b\right)=2g\left(b\right)/\zeta\left(b+1\right)$
is a proper PDF.
\end{singlespace}

Now evaluate (A.1) at $x=2,4$, giving
\[
\dfrac{2}{3}={\displaystyle \int_{0}^{\infty}}\left(2^{-b}\right)\dfrac{2g\left(b\right)}{\zeta\left(b+1\right)}db={\displaystyle \int_{0}^{\infty}}2^{-b}\eta\left(b\right)db\qquad\qquad\qquad\textrm{(A.3)}
\]
and
\[
\dfrac{2}{5}={\displaystyle \int_{0}^{\infty}}\left(4^{-b}\right)\dfrac{2}{\zeta\left(b+1\right)}g\left(b\right)db={\displaystyle \int_{0}^{\infty}}4^{-b}\eta\left(b\right)db,\qquad\qquad\qquad\textrm{(A.4)}
\]
respectively. Applying the Cauchy-Schwarz inequality to the integrals
in (A.2), (A.3), and (A.4) then yields
\[
\left({\displaystyle \int_{0}^{\infty}}\left(2^{-b}\right)\eta\left(b\right)db\right)^{2}\leq\left({\displaystyle \int_{0}^{\infty}}\left(1^{-b}\right)^{2}\eta\left(b\right)db\right)\left({\displaystyle \int_{0}^{\infty}}\left(2^{-b}\right)^{2}\eta\left(b\right)db\right)
\]
\[
\Longrightarrow\left(\dfrac{2}{3}\right)^{2}\leq\left(1\right)\left(\dfrac{2}{5}\right)\Longrightarrow\dfrac{4}{9}\leq\dfrac{2}{5},
\]
which is a clear contradiction. Thus, the posited mixing PDF $g\left(b\right)$
cannot exist.

The case of discrete mixtures can be handled in a parallel manner
by (i) assuming the existence of a mixing PMF $g\left(b\right)$ and
(ii) replacing all integrals with respect to $b\in\mathbb{R}_{>0}$
with summations over $b\in\mathbb{B}$ for some countable set $\mathbb{B}\subset\mathbb{R}_{>0}$.
\end{proof}
\begin{singlespace}
\noindent\textbf{Proof of Proposition 1.\medskip{}
}

\noindent Consider the $\nu^{\textrm{th}}$ factorial moment of $X\mid a\sim\textrm{Zeta Tail}\left(a\right)$,
for $\nu\in\mathbb{Z}_{\geq1}$:
\[
\textrm{E}_{X\mid a}^{\left(\textrm{ZT}\right)}\left[\left(X\right)_{\nu}\right]=\textrm{E}_{X\mid a}^{\left(\textrm{ZT}\right)}\left[X\left(X-1\right)\left(X-2\right)\cdots\left(X-\nu+1\right)\right]
\]
\[
=a{\displaystyle \sum_{x=\nu}^{\infty}}x\left(x-1\right)\left(x-2\right)\cdots\left(x-\nu+1\right)\zeta\left(x+1,a+1\right)
\]
\[
=a{\displaystyle \sum_{x=\nu}^{\infty}}\left[x\left(x-1\right)\left(x-2\right)\cdots\left(x-\nu+1\right)\sum_{i=0}^{\infty}\dfrac{1}{\left(i+a+1\right)^{x+1}}\right]
\]
\[
=a\sum_{i=0}^{\infty}\left[{\displaystyle \sum_{x=\nu}^{\infty}}\dfrac{x\left(x-1\right)\left(x-2\right)\cdots\left(x-\nu+1\right)}{\left(i+a+1\right)^{x+1}}\right]
\]
\[
=a\sum_{i=0}^{\infty}\left[\dfrac{1}{\left(i+a+1\right)^{\nu+1}}{\displaystyle \sum_{\tau=0}^{\infty}}\dfrac{\left(\tau+\nu\right)\left(\tau+\nu-1\right)\left(\tau+\nu-2\right)\cdots\left(\tau+1\right)}{\left(i+a+1\right)^{\tau}}\right]
\]
\[
=a\nu!\sum_{i=0}^{\infty}\left[\dfrac{1}{\left(i+a+1\right)^{\nu+1}}{\displaystyle \sum_{\tau=0}^{\infty}}\dbinom{\tau+\nu}{\tau}\dfrac{1}{\left(i+a+1\right)^{\tau}}\right]
\]
\[
=a\nu!\sum_{i=0}^{\infty}\left[\dfrac{1}{\left(i+a+1\right)^{\nu+1}}\left(1-\dfrac{1}{i+a+1}\right)^{-\left(\nu+1\right)}\right]
\]
\[
=a\nu!\sum_{i=0}^{\infty}\dfrac{1}{\left(i+a\right)^{\nu+1}}
\]
\[
=a\nu!\zeta\left(\nu+1,a\right).\qquad\qquad\qquad\textrm{(A.5)}
\]

\end{singlespace}

Now, since
\[
X^{\kappa}={\displaystyle \sum_{\nu=1}^{\kappa}}S\left(\kappa,\nu\right)\left(X\right)_{\nu}
\]
for $\kappa\in\mathbb{Z}_{\geq1}$, it follows from (A.5) that the
$\kappa^{\textrm{th}}$ raw moment of $X\mid a\sim\textrm{Zeta Tail}\left(a\right)$
is given by (7).
\begin{flushright}
$\boxempty$
\par\end{flushright}

\begin{singlespace}
\noindent\textbf{Proof of Proposition 2.\medskip{}
}
\end{singlespace}

\noindent (i) We begin with the probability generating function of
$X\mid a\sim\textrm{Zeta Tail}\left(a\right)$,
\[
\textrm{G}_{X\mid a}^{\left(\textrm{ZT}\right)}\left(z\right)=\textrm{E}_{X\mid a}^{\left(\textrm{ZT}\right)}\left[z^{X}\right]
\]
\[
=a{\displaystyle \sum_{x=1}^{\infty}}z^{x}\zeta\left(x+1,a+1\right)
\]
\[
=a{\displaystyle \sum_{x=1}^{\infty}}\left[z^{x}\sum_{i=0}^{\infty}\dfrac{1}{\left(i+a+1\right)^{x+1}}\right]
\]
\[
=a\sum_{i=0}^{\infty}\left[\dfrac{1}{i+a+1}{\displaystyle \sum_{x=1}^{\infty}}\left(\dfrac{z}{i+a+1}\right)^{x}\right]
\]
\[
=a\sum_{i=0}^{\infty}\left(\dfrac{1}{i+a+1}\right)\dfrac{z\left(i+a+1\right)^{-1}}{1-z\left(i+a+1\right)^{-1}}
\]
\[
=a\sum_{i=0}^{\infty}\dfrac{z}{\left(i+a+1\right)\left(i+a+1-z\right)}
\]
\[
=a\sum_{i=0}^{\infty}\left(\dfrac{1}{i+a+1-z}-\dfrac{1}{i+a+1}\right)
\]
\[
=a\left(\psi\left(a+1\right)-\psi\left(a+1-z\right)\right),\qquad\qquad\qquad\textrm{(A.6)}
\]
for $\left|z\right|<a+1$.

From (A.6), it is easy to solve for the associated moment-generating
function ($\textrm{M}_{X\mid a}^{\left(\textrm{ZT}\right)}\left(t\right)$),
Laplace transform ($\mathcal{L}_{X\mid a}^{\left(\textrm{ZT}\right)}\left(s\right)$),
and characteristic function ($\varphi_{X\mid a}^{\left(\textrm{ZT}\right)}\left(\omega\right)$)
as follows:

\begin{singlespace}
\noindent (ii) Setting $z=e^{t}$ implies
\[
\textrm{M}_{X\mid a}^{\left(\textrm{ZT}\right)}\left(t\right)=\textrm{E}_{X\mid a}^{\left(\textrm{ZT}\right)}\left[e^{tX}\right]
\]
\[
=a\left(\psi\left(a+1\right)-\psi\left(a+1-e^{t}\right)\right),
\]
for $t<\ln\left(a+1\right)$;
\end{singlespace}

\begin{singlespace}
\noindent (iii) Setting $z=e^{-s}$ implies
\end{singlespace}

\[
\mathcal{L}_{X\mid a}^{\left(\textrm{ZT}\right)}\left(s\right)=\textrm{E}_{X\mid a}^{\left(\textrm{ZT}\right)}\left[e^{-sX}\right]
\]
\[
=a\left(\psi\left(a+1\right)-\psi\left(a+1-e^{-s}\right)\right),
\]
for $s>-\ln\left(a+1\right)$; and

\begin{singlespace}
\noindent (iv) Setting $z=e^{i\omega}$ implies
\[
\varphi_{X\mid a}^{\left(\textrm{ZT}\right)}\left(\omega\right)=\textrm{E}_{X\mid a}^{\left(\textrm{ZT}\right)}\left[e^{i\omega X}\right]
\]
\[
=a\left(\psi\left(a+1\right)-\psi\left(a+1-e^{i\omega}\right)\right),
\]
for $\omega\in\mathbb{R}$.
\end{singlespace}

\begin{flushright}
$\boxempty$
\par\end{flushright}

\begin{singlespace}
\noindent\textbf{Proof of Proposition 3.\medskip{}
}
\end{singlespace}

\noindent (i) The hazard function is given by
\[
h_{X\mid a}^{\left(\textrm{ZT}\right)}\left(x\right)=\dfrac{f_{X\mid a}^{\left(\textrm{ZT}\right)}\left(x\right)}{1-F_{X\mid a}^{\left(\textrm{ZT}\right)}\left(x\right)}
\]
\[
=\dfrac{a\zeta\left(x+1,a+1\right)}{1-a{\displaystyle \sum_{k=1}^{x}}\zeta\left(k+1,a+1\right)}
\]
\[
=\dfrac{\zeta\left(x+1,a+1\right)}{{\displaystyle \sum_{k=x+1}^{\infty}}\zeta\left(k+1,a+1\right)},\qquad\qquad\qquad\textrm{(A.7)}
\]
for $x\in\mathbb{Z}_{\geq1}$. Now consider the ratio\newpage
\[
\dfrac{h_{X\mid a}^{\left(\textrm{ZT}\right)}\left(x+1\right)}{h_{X\mid a}^{\left(\textrm{ZT}\right)}\left(x\right)}=\dfrac{\zeta\left(x+2,a+1\right){\displaystyle \sum_{k=x+1}^{\infty}}\zeta\left(k+1,a+1\right)}{\zeta\left(x+1,a+1\right){\displaystyle \sum_{k=x+2}^{\infty}}\zeta\left(k+1,a+1\right)},
\]
which is less than $1$ for all $x$ if and only if the hazard function
is strictly decreasing in $x$. We then note that
\[
\dfrac{h_{X\mid a}^{\left(\textrm{ZT}\right)}\left(x+1\right)}{h_{X\mid a}^{\left(\textrm{ZT}\right)}\left(x\right)}<1
\]
\[
\Longleftrightarrow\dfrac{{\displaystyle \sum_{k=x+1}^{\infty}}\zeta\left(k+1,a+1\right)}{{\displaystyle \sum_{k=x+2}^{\infty}}\zeta\left(k+1,a+1\right)}<\dfrac{\zeta\left(x+1,a+1\right)}{\zeta\left(x+2,a+1\right)}
\]
\[
\Longleftrightarrow\dfrac{{\displaystyle \sum_{k=x+1}^{\infty}}\left[{\displaystyle \sum_{i=0}^{\infty}}\dfrac{1}{\left(i+a+1\right)^{k+1}}\right]}{{\displaystyle \sum_{k=x+2}^{\infty}}\left[{\displaystyle \sum_{i=0}^{\infty}}\dfrac{1}{\left(i+a+1\right)^{k+1}}\right]}<\dfrac{{\displaystyle \sum_{i=0}^{\infty}}\dfrac{1}{\left(i+a+1\right)^{x+1}}}{{\displaystyle \sum_{i=0}^{\infty}}\dfrac{1}{\left(i+a+1\right)^{x+2}}}
\]
\[
\Longleftrightarrow\dfrac{{\displaystyle \sum_{i=0}^{\infty}}\left[{\displaystyle \sum_{k=x+1}^{\infty}}\dfrac{1}{\left(i+a+1\right)^{k+1}}\right]}{{\displaystyle \sum_{i=0}^{\infty}}\left[{\displaystyle \sum_{k=x+2}^{\infty}}\dfrac{1}{\left(i+a+1\right)^{k+1}}\right]}<\dfrac{{\displaystyle \sum_{i=0}^{\infty}}\dfrac{1}{\left(i+a+1\right)^{x+1}}}{{\displaystyle \sum_{i=0}^{\infty}}\dfrac{1}{\left(i+a+1\right)^{x+2}}}
\]
\[
\Longleftrightarrow\dfrac{{\displaystyle \sum_{i=0}^{\infty}}\dfrac{\left(i+a+1\right)^{-\left(x+2\right)}}{1-\left(i+a+1\right)^{-1}}}{{\displaystyle \sum_{i=0}^{\infty}}\dfrac{\left(i+a+1\right)^{-\left(x+3\right)}}{1-\left(i+a+1\right)^{-1}}}<\dfrac{{\displaystyle \sum_{i=0}^{\infty}}\dfrac{1}{\left(i+a+1\right)^{x+1}}}{{\displaystyle \sum_{i=0}^{\infty}}\dfrac{1}{\left(i+a+1\right)^{x+2}}}
\]
\[
\Longleftrightarrow\dfrac{{\displaystyle \sum_{i=0}^{\infty}}\dfrac{\left(i+a+1\right)^{-\left(x+1\right)}}{i+a}}{{\displaystyle \sum_{i=0}^{\infty}}\dfrac{\left(i+a+1\right)^{-\left(x+2\right)}}{i+a}}<\dfrac{{\displaystyle \sum_{i=0}^{\infty}}\dfrac{1}{\left(i+a+1\right)^{x+1}}}{{\displaystyle \sum_{i=0}^{\infty}}\dfrac{1}{\left(i+a+1\right)^{x+2}}}
\]
\[
\Longleftrightarrow\dfrac{{\displaystyle \sum_{i=0}^{\infty}}\dfrac{\left(i+a\right)^{-1}}{\left(i+a+1\right)^{x+1}}}{{\displaystyle \sum_{i=0}^{\infty}}\dfrac{\left(i+a\right)^{-1}\left(i+a+1\right)^{-1}}{\left(i+a+1\right)^{x+1}}}<\dfrac{{\displaystyle \sum_{i=0}^{\infty}}\dfrac{1}{\left(i+a+1\right)^{x+1}}}{{\displaystyle \sum_{i=0}^{\infty}}\dfrac{\left(i+a+1\right)^{-1}}{\left(i+a+1\right)^{x+1}}}
\]
\[
\Longleftrightarrow\left({\displaystyle \sum_{i=0}^{\infty}}\dfrac{\left(i+a\right)^{-1}}{\left(i+a+1\right)^{x+1}}\right)\left({\displaystyle \sum_{i=0}^{\infty}}\dfrac{\left(i+a+1\right)^{-1}}{\left(i+a+1\right)^{x+1}}\right)
\]
\[
<\left({\displaystyle \sum_{i=0}^{\infty}}\dfrac{1}{\left(i+a+1\right)^{x+1}}\right)\left({\displaystyle \sum_{i=0}^{\infty}}\dfrac{\left(i+a\right)^{-1}\left(i+a+1\right)^{-1}}{\left(i+a+1\right)^{x+1}}\right)
\]
\[
\Longleftrightarrow\left(\sum_{i=0}^{\infty}\omega_{i}\dfrac{1}{i+a}\right)\left(\sum_{i=0}^{\infty}\omega_{i}\dfrac{1}{i+a+1}\right)<\sum_{i=0}^{\infty}\omega_{i}\left(\dfrac{1}{i+a}\right)\left(\dfrac{1}{i+a+1}\right),\qquad\textrm{(A.8)}
\]
where
\[
\omega_{i}=\dfrac{\dfrac{1}{\left(i+a+1\right)^{x+1}}}{{\displaystyle \sum_{\tau=0}^{\infty}}\dfrac{1}{\left(\tau+a+1\right)^{x+1}}},\quad\sum_{i=0}^{\infty}\omega_{i}=1,
\]
and (A.8) follows from Chebyshev's sum inequality.

The expression in (A.7) further implies
\[
h_{X\mid a}^{\left(\textrm{ZT}\right)}\left(x\right)=\dfrac{{\displaystyle \sum_{i=0}^{\infty}}\dfrac{1}{\left(i+a+1\right)^{x+1}}}{{\displaystyle \sum_{k=x+1}^{\infty}}\left[{\displaystyle \sum_{i=0}^{\infty}}\dfrac{1}{\left(i+a+1\right)^{k+1}}\right]}
\]
\[
=\dfrac{{\displaystyle \sum_{i=0}^{\infty}}\dfrac{1}{\left(i+a+1\right)^{x+1}}}{{\displaystyle \sum_{i=0}^{\infty}}\left[{\displaystyle \sum_{k=x+1}^{\infty}}\dfrac{1}{\left(i+a+1\right)^{k+1}}\right]}
\]
\[
=\dfrac{{\displaystyle \sum_{i=0}^{\infty}}\dfrac{1}{\left(i+a+1\right)^{x+1}}}{{\displaystyle \sum_{i=0}^{\infty}}\dfrac{\left(i+a+1\right)^{-\left(x+2\right)}}{1-\left(i+a+1\right)^{-1}}}
\]
\[
=\dfrac{{\displaystyle \sum_{i=0}^{\infty}}\dfrac{1}{\left(i+a+1\right)^{x+1}}}{{\displaystyle \sum_{i=0}^{\infty}}\dfrac{1}{\left(i+a\right)\left(i+a+1\right)^{x+1}}}
\]
\[
=\dfrac{\dfrac{1}{\left(a+1\right)^{x+1}}+{\displaystyle \sum_{i=1}^{\infty}}\dfrac{1}{\left(i+a+1\right)^{x+1}}}{\dfrac{1}{a\left(a+1\right)^{x+1}}+{\displaystyle \sum_{i=1}^{\infty}}\dfrac{1}{\left(i+a\right)\left(i+a+1\right)^{x+1}}}
\]
\[
=\dfrac{a+a{\displaystyle \sum_{i=1}^{\infty}}\left(\dfrac{a+1}{i+a+1}\right)^{x+1}}{1+a{\displaystyle \sum_{i=1}^{\infty}\dfrac{1}{\left(i+a\right)}}\left(\dfrac{a+1}{i+a+1}\right)^{x+1}},
\]
and it follows (from dominated convergence of the two summations)
that
\[
h_{X\mid a}^{\left(\textrm{ZT}\right)}\left(x\right)=\dfrac{a+a\left(\dfrac{a+1}{a+2}\right)^{x+1}\left(1+o\left(1\right)\right)}{1+\left(\dfrac{a}{a+1}\right)\left(\dfrac{a+1}{a+2}\right)^{x+1}\left(1+o\left(1\right)\right)}
\]
\[
=\dfrac{a+O\left(\left(\dfrac{a+1}{a+2}\right)^{x+1}\right)}{1+O\left(\left(\dfrac{a+1}{a+2}\right)^{x+1}\right)}
\]
 for large $x$. Thus,
\[
\underset{x\rightarrow\infty}{\lim}h_{X\mid a}^{\left(\textrm{ZT}\right)}\left(x\right)=a.
\]

\begin{singlespace}
\noindent (ii) The mean residual-life function is given by
\[
e_{X\mid a}^{\left(\textrm{ZT}\right)}\left(x\right)=\textrm{E}_{X\mid a}^{\left(\textrm{ZT}\right)}\left[\left.X-x\right|X>x\right]
\]
\[
={\displaystyle \sum_{k=x+1}^{\infty}}\left(k-x\right)\dfrac{\zeta\left(k+1,a+1\right)}{{\displaystyle \sum_{\tau=x+1}^{\infty}}\zeta\left(\tau+1,a+1\right)}
\]
\[
=\dfrac{{\displaystyle \sum_{k=1}^{\infty}}\left(k-x\right)\zeta\left(k+1,a+1\right)-{\displaystyle \sum_{k=1}^{x-1}}\left(k-x\right)\zeta\left(k+1,a+1\right)}{{\displaystyle \sum_{\tau=x+1}^{\infty}}\zeta\left(\tau+1,a+1\right)}
\]
\[
=\dfrac{\dfrac{\textrm{E}_{X\mid a}^{\left(\textrm{ZT}\right)}\left[X\right]-x}{a}-{\displaystyle \sum_{k=1}^{x-1}}\left(k-x\right)\zeta\left(k+1,a+1\right)}{{\displaystyle \sum_{\tau=x+1}^{\infty}}\zeta\left(\tau+1,a+1\right)}
\]
\[
=\dfrac{\zeta\left(2,a\right)-\dfrac{x}{a}+{\displaystyle \sum_{k=1}^{x-1}}\left(x-k\right)\zeta\left(k+1,a+1\right)}{{\displaystyle \sum_{k=x+1}^{\infty}}\zeta\left(k+1,a+1\right)}\qquad\qquad\textrm{(A.9)}
\]
for $x\in\mathbb{Z}_{\geq1}$. Then, since $h_{X\mid a}^{\left(\textrm{ZT}\right)}\left(x\right)$
is strictly decreasing in $x$ (as shown in part (i)), it follows
that 
\[
\overline{h}_{X\mid a}^{\left(\textrm{ZT}\right)}\left(x\right)=\dfrac{f_{X\mid a}^{\left(\textrm{ZT}\right)}\left(x\right)}{1-F_{X\mid a}^{\left(\textrm{ZT}\right)}\left(x-1\right)}
\]
\[
=\dfrac{h_{X\mid a}^{\left(\textrm{ZT}\right)}\left(x\right)}{1+h_{X\mid a}^{\left(\textrm{ZT}\right)}\left(x\right)}
\]
is also strictly decreasing, which in turn implies that $e_{X\mid a}^{\left(\textrm{ZT}\right)}\left(x\right)$
must be strictly increasing (see, e.g., p. 619 of Hollander and Proschan,
1984).
\end{singlespace}

Furthermore, (A.9) implies
\[
e_{X\mid a}^{\left(\textrm{ZT}\right)}\left(x\right)=\dfrac{{\displaystyle \sum_{i=0}^{\infty}}\dfrac{1}{\left(i+a\right)^{2}}-\dfrac{x}{a}+{\displaystyle \sum_{k=1}^{x-1}}\left(x-k\right){\displaystyle \sum_{i=0}^{\infty}}\dfrac{1}{\left(i+a+1\right)^{k+1}}}{{\displaystyle \sum_{k=x+1}^{\infty}}\left[{\displaystyle \sum_{i=0}^{\infty}}\dfrac{1}{\left(i+a+1\right)^{k+1}}\right]}
\]
\[
=\dfrac{{\displaystyle \sum_{i=0}^{\infty}}\dfrac{1}{\left(i+a\right)^{2}}-\dfrac{x}{a}+{\displaystyle \sum_{i=0}^{\infty}}\left[{\displaystyle \sum_{k=1}^{x-1}}\dfrac{\left(x-k\right)}{\left(i+a+1\right)^{k+1}}\right]}{{\displaystyle \sum_{i=0}^{\infty}}\left[{\displaystyle \sum_{k=x+1}^{\infty}}\dfrac{1}{\left(i+a+1\right)^{k+1}}\right]}
\]
\[
=\dfrac{{\displaystyle \sum_{i=0}^{\infty}}\dfrac{1}{\left(i+a\right)^{2}}-\dfrac{x}{a}+{\displaystyle \sum_{i=0}^{\infty}}\left[{\displaystyle \sum_{k=1}^{x-1}}\dfrac{\left(x-k\right)}{\left(i+a+1\right)^{k+1}}\right]}{{\displaystyle \sum_{i=0}^{\infty}}\dfrac{\left(i+a+1\right)^{-\left(x+2\right)}}{1-\left(i+a+1\right)^{-1}}}
\]
\[
=\dfrac{{\displaystyle \sum_{i=0}^{\infty}}\dfrac{1}{\left(i+a\right)^{2}}-\dfrac{x}{a}+{\displaystyle \sum_{i=0}^{\infty}}\left[{\displaystyle \sum_{k=1}^{x-1}}\dfrac{\left(x-k\right)}{\left(i+a+1\right)^{k+1}}\right]}{{\displaystyle \sum_{i=0}^{\infty}}\dfrac{1}{\left(i+a\right)\left(i+a+1\right)^{x+1}}}
\]
\[
=\dfrac{{\displaystyle \sum_{i=0}^{\infty}}\dfrac{1}{\left(i+a\right)^{2}}-\dfrac{x}{a}+{\displaystyle \sum_{i=0}^{\infty}}\left[\dfrac{x}{\left(i+a\right)\left(i+a+1\right)}+\dfrac{\left(i+a+1\right)^{-x}-1}{\left(i+a\right)^{2}}\right]}{{\displaystyle \sum_{i=0}^{\infty}}\dfrac{1}{\left(i+a\right)\left(i+a+1\right)^{x+1}}}
\]
\[
=\dfrac{-\dfrac{x}{a}+\dfrac{x}{a}+{\displaystyle \sum_{i=0}^{\infty}}\dfrac{1}{\left(i+a\right)^{2}\left(i+a+1\right)^{x}}}{{\displaystyle \sum_{i=0}^{\infty}}\dfrac{1}{\left(i+a\right)\left(i+a+1\right)^{x+1}}}
\]
\[
=\dfrac{\dfrac{1}{a^{2}\left(a+1\right)^{x}}+{\displaystyle \sum_{i=1}^{\infty}}\dfrac{1}{\left(i+a\right)^{2}\left(i+a+1\right)^{x}}}{\dfrac{1}{a\left(a+1\right)^{x+1}}+{\displaystyle \sum_{i=1}^{\infty}}\dfrac{1}{\left(i+a\right)\left(i+a+1\right)^{x+1}}}
\]
\[
=\dfrac{a+1+a^{2}{\displaystyle \sum_{i=1}^{\infty}}\dfrac{\left(a+1\right)}{\left(i+a\right)^{2}}\left(\dfrac{a+1}{i+a+1}\right)^{x}}{a+a^{2}{\displaystyle \sum_{i=1}^{\infty}\dfrac{1}{i+a}\left(\dfrac{a+1}{i+a+1}\right)^{x+1}}},
\]
and one can see (by dominated convergence of the two summations) that
\[
e_{X\mid a}^{\left(\textrm{ZT}\right)}\left(x\right)=\dfrac{a+1+\dfrac{a^{2}\left(a+1\right)}{\left(a+1\right)^{2}}\left(\dfrac{a+1}{a+2}\right)^{x}\left(1+o\left(1\right)\right)}{a+\dfrac{a^{2}}{a+1}\left(\dfrac{a+1}{a+2}\right)^{x+1}\left(1+o\left(1\right)\right)}
\]
\[
=\dfrac{a+1+O\left(\left(\dfrac{a+1}{a+2}\right)^{x}\right)}{a+O\left(\left(\dfrac{a+1}{a+2}\right)^{x+1}\right)}
\]
for large $x$. Thus,
\[
\underset{x\rightarrow\infty}{\lim}e_{X\mid a}^{\left(\textrm{ZT}\right)}\left(x\right)=\dfrac{a+1}{a}.
\]

\begin{flushright}
$\boxempty$
\par\end{flushright}

\begin{singlespace}
\noindent\textbf{Proof of Proposition 4.\medskip{}
}

\noindent In (5), the $\textrm{Zeta Tail}\left(a=1\right)$ PMF was
written as a countable mixture of $\textrm{Geometric}\left(p\right)$
PMFs. This formulation can be extended to $a\in\mathbb{R}_{>0}$ by
taking $p=\left(m+a-1\right)/\left(m+a\right)$ and replacing the
$\textrm{Quadratic}\left(c=1\right)$ mixing distribution with the
$\textrm{Quadratic}\left(c=a\right)$ distribution, characterized
by\newpage
\[
f_{X\mid c=a}^{\left(\textrm{Q}\right)}\left(m\right)=\left.\dfrac{c}{\left(m-1+c\right)\left(m+c\right)}\right|_{c=a},\quad m\in\mathbb{Z}_{\geq1}
\]
\[
=\dfrac{a}{\left(m+a-1\right)\left(m+a\right)}.
\]
Then
\[
\sum_{m=1}^{\infty}f_{X\mid p=\tfrac{m+a-1}{m+a}}^{\left(\textrm{G}\right)}\left(x\right)f_{X\mid c=a}^{\left(\textrm{Q}\right)}\left(m\right)=\sum_{m=1}^{\infty}\left(\dfrac{m+a-1}{m+a}\right)\left(\dfrac{1}{m+a}\right)^{x-1}\dfrac{a}{\left(m+a-1\right)\left(m+a\right)}
\]
\[
=\sum_{m=1}^{\infty}a\left(\dfrac{1}{m+a}\right)^{x+1}
\]
\[
=a\sum_{m=0}^{\infty}\dfrac{1}{\left(m+a+1\right)^{x+1}}
\]
\[
=a\zeta\left(x+1,a+1\right)=f_{X\mid a}^{\left(\textrm{ZT}\right)}\left(x\right).
\]

\end{singlespace}

\begin{flushright}
$\boxempty$
\par\end{flushright}

\begin{thebibliography}{10}
\begin{singlespace}
\bibitem{key-1}Chiu, S. N. and Yin, C., 2014, ``On the Complete
Monotonicity of the Compound Geometric Convolution with Applications
in Risk Theory'', \emph{Scandinavian Actuarial Journal}, 2014, 2,
116-124.

\bibitem{key-2}Daniels, M. J. and Hogan, J. W., 2008, \emph{Missing
Data in Longitudinal Studies: Strategies for Bayesian Modeling and
Sensitivity Analysis}, Chapman and Hall, London, UK.

\bibitem{key-3}Deni, S. M. and Jemain, A. A., 2009, ``Fitting the
Distribution of Dry and Wet Spells with Alternative Probability Models'',
\emph{Meteorology and Atmospheric Physics}, 104, 13-27.

\bibitem{key-4}Dobi-Wantuch, J., Mika, J., and Szeidl, L., 2000,
``Modelling Wet and Dry Spells with Mixture Distributions'', \emph{Meteorology
and Atmospheric Physics}, 73, 245-256.

\bibitem{key-5}Foufoula-Georgiou, E. and Lettenmaier, D. P., 1987,
``A Markov Renewal Model for Rainfall Occurrences'', \emph{Water
Resources Research}, 23, 5, 875-884.

\bibitem{key-6}Hesselager, O., 1994, ``A Recursive Procedure for
Calculation of Some Compound Distributions'', \emph{ASTIN Bulletin},
24, 1, 19-32.

\bibitem{key-7}Hollander, M. and Proschan, F., 1984, ``Nonparametric
Concepts and Methods in Reliability'', 613-655, in P. R. Krishnaiah
and P. K. Sen, eds., \emph{Handbook of Statistics, Volume 4}, Elsevier
B.V., Amsterdam.

\bibitem{key-8}Hu, C.-Y., Iksanov, A. M., Lin, G. D., and Zakusylo,
O. K., 2006, ``The Hurwitz Zeta Distribution'', \emph{Australian
and New Zealand Journal of Statistics}, 48, 1, 1-6.

\bibitem{key-9}Keith, W. J., 2010, ``Sequences of Density $\zeta\left(K\right)-1$'',
\emph{Integers}, 10, 233-241.

\bibitem{key-10}Lee, C.-E. and Kim, S. U., 2017, ``Applicability
of Zero-Inflated Models to Fit the Torrential Rainfall Count Data
with Extra Zeros in South Korea'', \emph{Water}, 9, 123.

\bibitem{key-11}Lüxmann-Ellinghaus, U., 1987, ``On the Identifiability
of Mixtures of Infinitely Divisible Power-Series Distributions'',
\emph{Statistics and Probability Letters}, 5, 375-378.

\bibitem{key-12}Panjer, H. H., 1981, ``Recursive Evaluation of a
Family of Compound Distributions'', \emph{ASTIN Bulletin}, 12, 1,
22-26.

\bibitem{key-13}Powers, M. R. and Xu, J., 2025a, ``A Versatility
Measure for Parametric Risk Models'', \emph{Risk Sciences}, 1, 100017.

\bibitem{key-14}Powers, M. R. and Xu, J., 2025b, ``A Criterion for
Extending Continuous-Mixture Identifiability Results'', \emph{Journal
of Statistical Theory and Applications}, 24, 515-533.

\bibitem{key-15}Powers, M. R. and Xu, J., 2025c, ``Assessing Risk
Heterogeneity through Heavy-Tailed Frequency and Severity Mixtures'',
arXiv:2505.04795.

\bibitem{key-16}Tzaninis, S. M. and Bozikas, A., 2026, ``Extensions
of Panjer\textquoteright s Recursion for Mixed Compound Distributions'',
\emph{Journal of Computational and Applied Mathematics}, 476, 117138.

\bibitem{key-17}Willmot, G. E. and Woo, J.-K., 2013, ``Some Distributional
Properties of a Class of Counting Distributions with Claims Analysis
Applications'', \emph{ASTIN Bulletin}, 43, 2, 189-212.
\end{singlespace}

\end{thebibliography}
\end{document}